\documentclass[10pt,conference]{IEEEtran}

\usepackage{subfigure}
\usepackage{booktabs}
\usepackage{graphicx, epsfig}
\usepackage{tikz}
\usepackage{color}

\usepackage{pseudocode}
\usepackage{blindtext}
\usepackage{listings}
\usepackage[english]{babel}
\usepackage{xurl}

\usepackage{amsmath}
\usepackage{libertine}

\definecolor{dkgreen}{rgb}{0,0.6,0}
\definecolor{gray}{rgb}{0.5,0.5,0.5}
\definecolor{mauve}{rgb}{0.58,0,0.82}

\newcommand{\oldstuff}[1]{}
\newcommand{\old}[1]{}
\newcommand{\optional}[1]{}
\newcommand{\consider}[1]{}
\newcommand{\moved}[1]{}
\newcommand{\comments}[1]{}

\newcommand{\eg}{{e.g., }}

\newcommand{\ie}{{i.e., }}

\ifdefined\etal
\else
  \newcommand{\etal}{{et al. }} 
\fi

\newcommand{\note}[1]{{}}

\newenvironment{bullets}%
{\begin{list}{$\bullet$}{\setlength{\leftmargin}{1.5ex}
\setlength{\itemindent}{.5ex}}}
{\end{list}}

\newcommand{\bi}{\begin{itemize}}
\newcommand{\ei}{\end{itemize}}
\newcommand{\be}{\begin{enumerate}}
\newcommand{\ee}{\end{enumerate}}
\newcommand{\bb}{\begin{bullets}}
\newcommand{\eb}{\end{bullets}}

\begin{document}

\title{Adaptive Distributed Filtering of \\DDoS Traffic on the Internet}

\author{\IEEEauthorblockN{
    Jun Li\IEEEauthorrefmark{1},
    Devkishen Sisodia\IEEEauthorrefmark{1},
    Yebo Feng\IEEEauthorrefmark{1},
    Lumin Shi\IEEEauthorrefmark{1},
    Mingwei Zhang\IEEEauthorrefmark{1},
    Christopher Early\IEEEauthorrefmark{1},
    Peter Reiher\IEEEauthorrefmark{2}}
    \IEEEauthorblockA{
        \textit{\IEEEauthorrefmark{1}University of Oregon,
        \IEEEauthorrefmark{2}UCLA}\\
    Email: lijun@cs.uoregon.edu}
}
\maketitle

\begin{abstract}

Despite the proliferation of traffic filtering capabilities throughout the Internet,
attackers continue to launch distributed denial-of-service (DDoS) attacks to successfully overwhelm the victims with DDoS traffic.
In this paper, we introduce a distributed filtering system
that leverages nodes distributed along the paths of DDoS traffic to filter the DDoS traffic.
In particular, we focus on adaptive distributed filtering, a new direction in filtering DDoS traffic.
In our design, a subscriber to the distributed filtering service can act on behalf of a DDoS victim
and generate filtering rules that not only adapt to the most suitable and effective filtering granularity 
(\eg IP source address and a port number vs. an individual IP address vs. IP prefixes at different lengths),
but also adapt to the preferences of the subscriber
(\eg maximum coverage of DDoS traffic vs.  minimum collateral damage from dropping legitimate traffic vs.
minimum number of rules).

We design an efficient algorithm that can generate rules adaptive toward filtering granularities and objectives,
which can further help determine where to deploy generated rules for the best efficacy. 
We evaluated our system through both large-scale simulations based on real-world DDoS attack traces and pilot studies.
Our evaluations confirm that our algorithm can generate rules that adapt to every distinct filtering objective and achieve optimal results.
We studied the success rate and distribution of rule deployment under different Internet-scale rule deployment profiles, and
found a small number of autonomous systems can contribute disproportionately to the defense.
Our pilot studies also show
our adaptive distributed filtering system can
effectively defend against real-world DDoS attack traces in real time. 


\end{abstract}

\section{Introduction}
\label{sec:introduction}

Despite years of research and industry efforts that have led to a myriad of defense approaches,
the Internet continues to be severely susceptible to distributed denial-of-service (DDoS) attacks
and see DDoS attacks increase in both the amount and scale~\cite{akamai2020report}.
Among the most common DDoS attacks are high-volume DDoS
that overwhelm a victim's bandwidth,
in which such attacks can reach as high as
1.2 Tbps~\cite{Dyn2016},
1.35 Tbps~\cite{Github2018},
2.4 Tbps~\cite{microsoft2021attack},
or even 3.47 Tbps~\cite{microsoft2022attack},
with largest ever recorded packet per second-based DDoS at 809 Mpps~\cite{akamai2020ddos}.

On the other hand, the Internet has seen a proliferation of filtering capabilities throughout.
With Access Control Lists (ACLs) built into routers by vendors from day one,
broader usage of Border Gateway Protocol (BGP) flow specification~\cite{rfc5575},
the increasing deployment of software-defined networking (SDN), and so on,
Internet service providers (ISPs) at the core of the Internet or proxies and firewalls at the edge
are equipped and ready to filter traffic, including DDoS packets.
Indeed, many ISPs already perform various traffic filtering operations, including filtering DDoS traffic.

Meanwhile, DDoS filtering has often been at a single point at the victim end~\cite{arboraps},
not leveraging the ever-growing filtering capabilities inherent in the Internet.
Or, if it is conducted in multiple locations over the Internet,
little has been done to investigate how filtering may be adaptive to a multitude of orthogonal or conflicting factors,
such as filtering granularity 
(\eg IP source address and a port number vs. an individual IP address vs. IP prefixes at different lengths),
the tradeoff between DDoS traffic coverage, collateral damage from dropping legitimate traffic, and the number of DDoS-filtering rules,
and DDoS traffic with spoofed source IP addresses.

In this paper, we introduce adaptive distributed filtering.
We allow a subscriber who acts on behalf of a DDoS victim
to adaptively generate rules at the proper granularity and then
deploy them at the most suitable filtering nodes on the paths of DDoS traffic.
For example, although it is probably suitable to filter traffic from an entire IP prefix when a victim is under a severe DDoS attack,
if the volume of DDoS traffic from the prefix is low
and the volume of legitimate traffic from the prefix is high, 
it may be more preferable to filter traffic from individual IP addresses instead. 
Or, if there are two traffic flows appearing from the same IP address,
one of which is benign traffic and the other is DDoS traffic spoofing the source,
filtering traffic from the IP address becomes a dilemma
unless it happens only on the paths of the DDoS flow,
or the filtering must use both IP address and port numbers if the two flows share the same path but use different port numbers. 

Adaptive distributed filtering is challenging 
because there can be thousands of DDoS flows at prefix level,
millions of DDoS flows at IP address level,
and potentially even more at IP address plus port number level.
It will be prohibitively expensive to monitor traffic for every granularity and then determine filtering rules accordingly.
Furthermore, every victim may have their own objectives regarding
maximum coverage of DDoS traffic, minimum collateral damage to legitimate traffic, and minimum number of rules, 
while in general one cannot accomplish all objectives simultaneously.

We therefore develop an efficient rule-generation algorithm
that can not only generate rules with different granularities toward different objectives,
but also help determine where to deploy generated rules for the best efficacy.
We then evaluate our system that embraces the algorithm.
We first use large-scale simulations based on real-world DDoS attack traces to 
study the efficacy of rules generated,
then study their deployment success rate under different distributed Internet-scale filtering profiles,
and also experiment the efficacy and scalability of the entire system for DDoS mitigation
in real time against real-world DDoS attack traces.

The rest of this paper is organized as follows.
We first summarize related work in Section~\ref{sec:related}.
We then describe the adaptive distributed filtering model for DDoS mitigation in Section~\ref{sec:model} and
elaborate our system design in Section~\ref{sec:system}, 
followed by the implementation in Section~\ref{sec:implementation}.
We detail our results from evaluating our solution through simulations and pilot studies in Section~\ref{sec:evaluation} and
conclude the paper in Section~\ref{sec:conclusion}.
\section{Related Work}
\label{sec:related}

Although starting from day one
DDoS traffic was launched from all over the Internet toward a victim,
filtering DDoS traffic was initially done only at a single point at the victim end.
Network operators can install systems at a victim 
such as Arbor APS~\cite{arboraps} and FastNetMon~\cite{fastnetmon} 
to detect and filter DDoS traffic upon their arrival,
or manually connect (\eg via \textit{ssh}) to local routers or firewalls to install Access Control Lists (ACLs) to filter the DDoS traffic.
While victim-end solutions are relatively easy to implement and deploy,
they can incur a high defense cost due to resource requirements 
in terms of network connection and devices,
and often fail to mitigate attacks when a victim's inbound links are already inundated with DDoS traffic.

Filtering of DDoS traffic has thus evolved in two complementary directions.
One direction is to enhance the ability to filter DDoS traffic of a single unit,
whether it is a programmable switch that is powerful for detecting and mitigating DDoS traffic~\cite{Posidon,liu2021jaqen},
an ISP that can orchestrate multiple virtual machines to process and filter DDoS traffic (\cite{bohatei}),
or a DDoS-scrubbing service that can scrub DDoS traffic and forward only cleansed traffic to clients (\cite{akamaiscrubbing,liu2016middlepolice}).
The other direction, which is more related to this work, is to filter DDoS traffic at multiple locations, \ie distributed filtering,
which we further elaborate below. 

The most straightforward distributed filtering is probably filtering DDoS at their sources.
Example solutions include D-WARD~\cite{mirkovic2002attacking}, which installs rate-limiting filters at border routers in source networks,
and COSSACK~\cite{papadopoulos2003}, which deploys countermeasures at the ASes of attacking sources.
Such approaches face severe difficulties in locating a large number of DDoS sources and deploying filters against them.

There are also numerous methods that filter DDoS traffic at multiple upstream ISPs,
such as PushBack~\cite{mahajan2002pushback}, TVA~\cite{yang2005TVA}, AITF~\cite{Argyraki2005}, DefCOM~\cite{oikonomou2006framework}, 
StopIt~\cite{liu2008filter}, and RAD~\cite{kline2009rad}.
The foci of these methods are mostly to strengthen the capabilities of routers
in identifying, processing, and filtering DDoS packets,
including through the development of new communication protocols.
The advent of software-defined networking (SDN), due to its friendliness to deploying rules to filter traffic,
has also led to a variety of SDN-based DDoS filtering solutions
that allow a victim to express to filtering entities inside the network
their preferred traffic control policies~\cite{liu2016middlepolice},
specific DDoS-filtering requests~\cite{ramanathan2018senss}, or
blackholing rules~\cite{dietzel2018stellar}. 
Nonetheless, much is yet to be done to investigate how filtering may be adaptive to a multitude of situations,
including how filtering may use a dynamic granularity to most effectively filter DDoS traffic,
even when some DDoS traffic uses spoofed IP addresses,
how filtering rules may adapt to the preferred filtering objectives of users, and
how to achieve the best tradeoff between filtering DDoS traffic and avoiding the collateral damage to legitimate traffic without using too many rules.

\section{Adaptive Distributed Filtering Model}
\label{sec:model}

\subsection{Distributed Filtering}

While recent DDoS mitigation research investigated the mitigation of DDoS traffic 
at a programmable switch (\cite{Posidon,liu2021jaqen}) or an ISP (\cite{bohatei}),
we focus on a \textbf{distributed filtering model} where the mitigation happens in multiple different locations.
As a DDoS attack is to launch DDoS traffic from DDoS bots throughout the Internet towards a victim along many different paths,
the DDoS traffic can be filtered along these paths before they reach the victim,
so long as on the paths there are nodes that are set to help filter DDoS traffic and know what traffic are DDoS traffic to filter.
An exemplary distributed filtering approach is AITF (\cite{Argyraki2005}),
which filters DDoS traffic based on individual IP addresses of DDoS bots as close to the source as possible.

Our study is centered on having proper nodes on DDoS paths 
to deploy proper rules regarding what traffic is DDoS traffic and thus should be filtered,
including adaptively determining the \emph{proper} nodes and \emph{proper} rules.
We elaborate our design's fundamental differences from existing work, including AITF, in Section~\ref{sec:model.adaptive}.

The development of Internet has made distributed filtering of DDoS possible.
Nodes on the Internet, whether they are routers, programmable switches, middleboxes, or end-hosts,
are witnessing increased capabilities of inspecting and filtering network traffic.
A ``node'' can also be an ISP, as some ISPs on the Internet can provide DDoS-filtering service,
or a specialized DDoS-scrubbing service,
which typically consists of multiple geo-distributed data centers to filter traffic. 

Distributed filtering of DDoS is also often necessary.
If only filtering all the DDoS traffic at a single point, 
it has to be done at a point where most, if not all, of the DDoS traffic converges.
This point of filtering can be a DDoS victim server itself or the firewall of the DDoS victim's network, 
which all the DDoS traffic to the victim will hit.
The point can also be at the ISP of the victim's network, which may or may not see all the DDoS traffic to the victim,
dependent on whether the victim's network is solely dependent on the ISP for Internet connectivity, \ie single-homed, or multi-homed.
A clear disadvantage here is that at such a point the volume of DDoS traffic can be already too overwhelming to handle.
Another critical disadvantage, albeit more subtle, is its more severe susceptibility to IP spoofing:
Say a DDoS bot is spoofing the IP address of a benign end-host which is also sending traffic to the victim.
If filtering DDoS traffic using the source IP address at a single point where traffic converges,
it will also drop the legitimate traffic from the benign host.
On the other hand, distributed filtering will be less subject to IP spoofing, 
as we will elaborate later in Section~\ref{sec:model.adaptive}.

\begin{figure}[th!]
	\centering
	\subfigure[Internet scale] {
	  \includegraphics[width=0.33\textwidth]{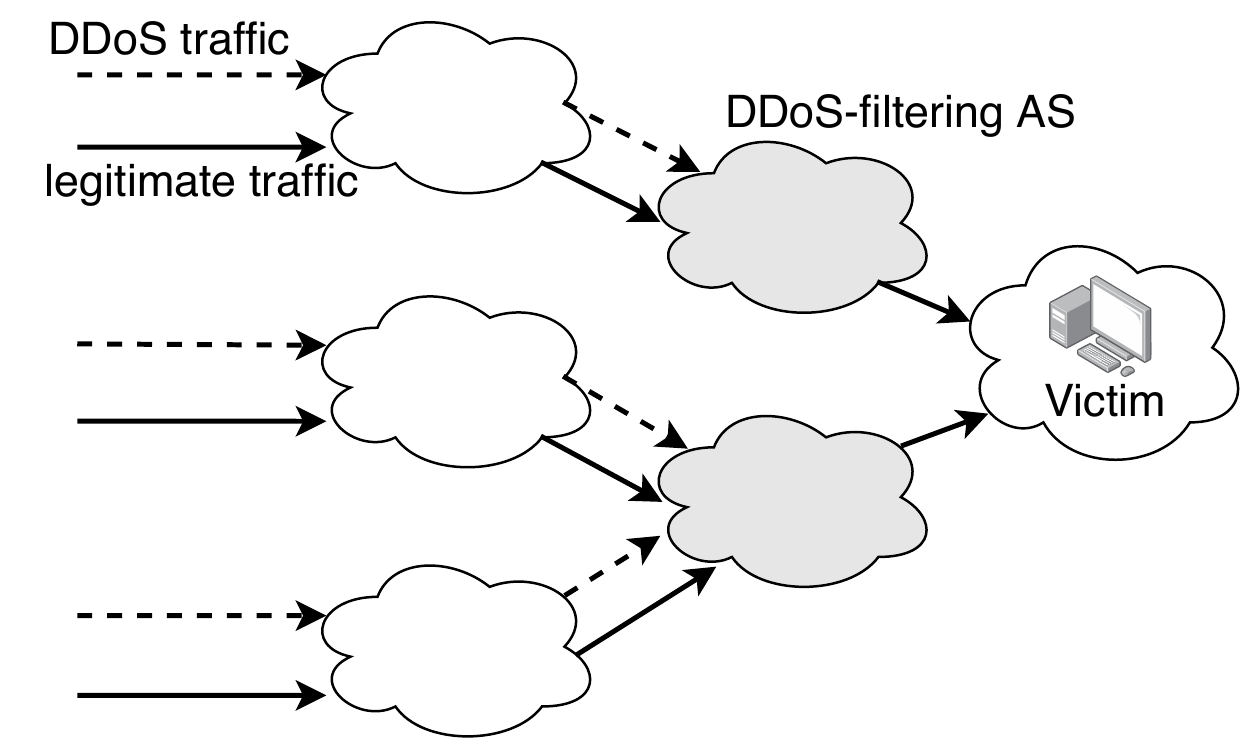}\label{fig:filter_as}
	  }
	\subfigure[ISP scale] {
	  \includegraphics[width=0.33\textwidth]{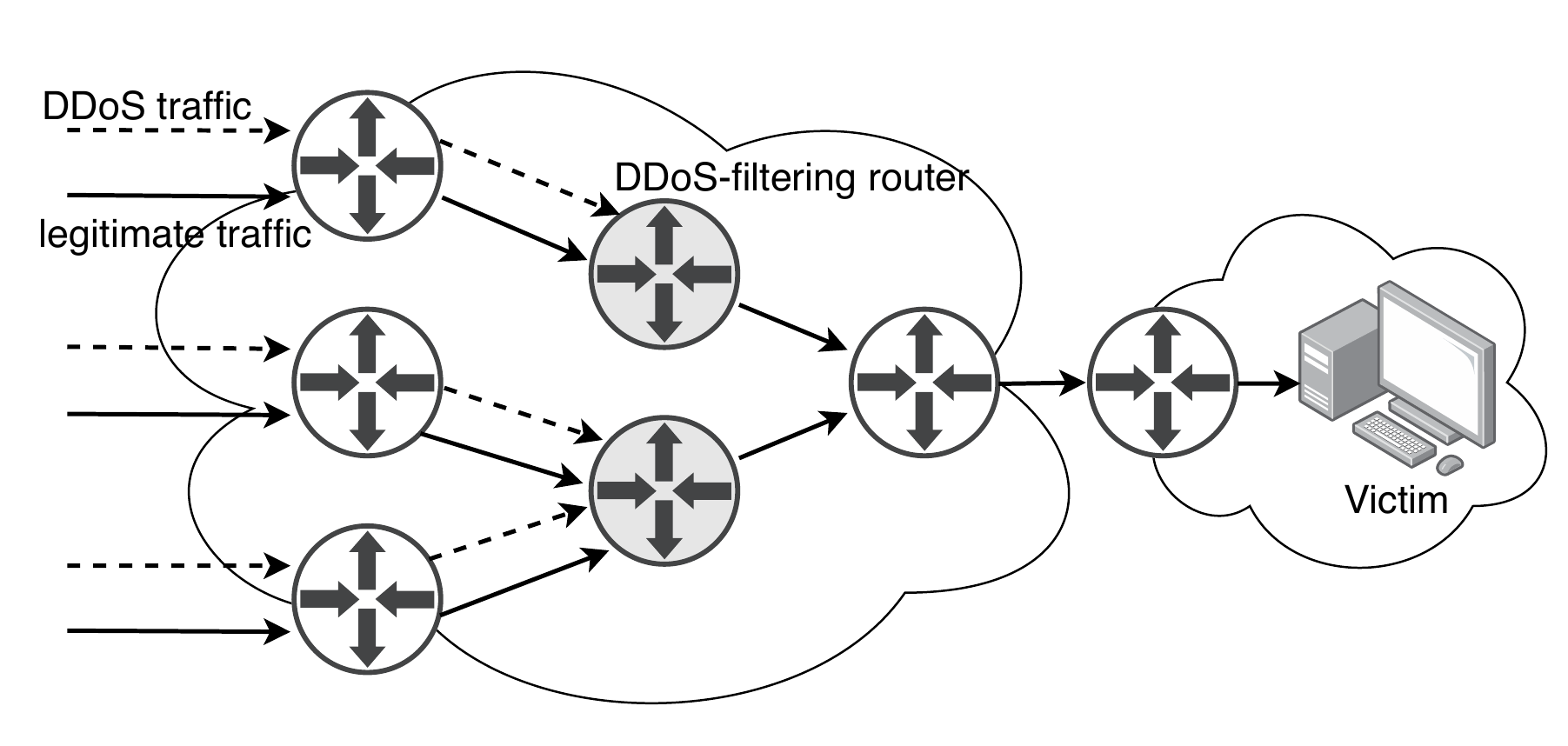}\label{fig:filter_router} 
	  }
	\subfigure[Scrubbing service scale] {
	  \includegraphics[width=0.33\textwidth]{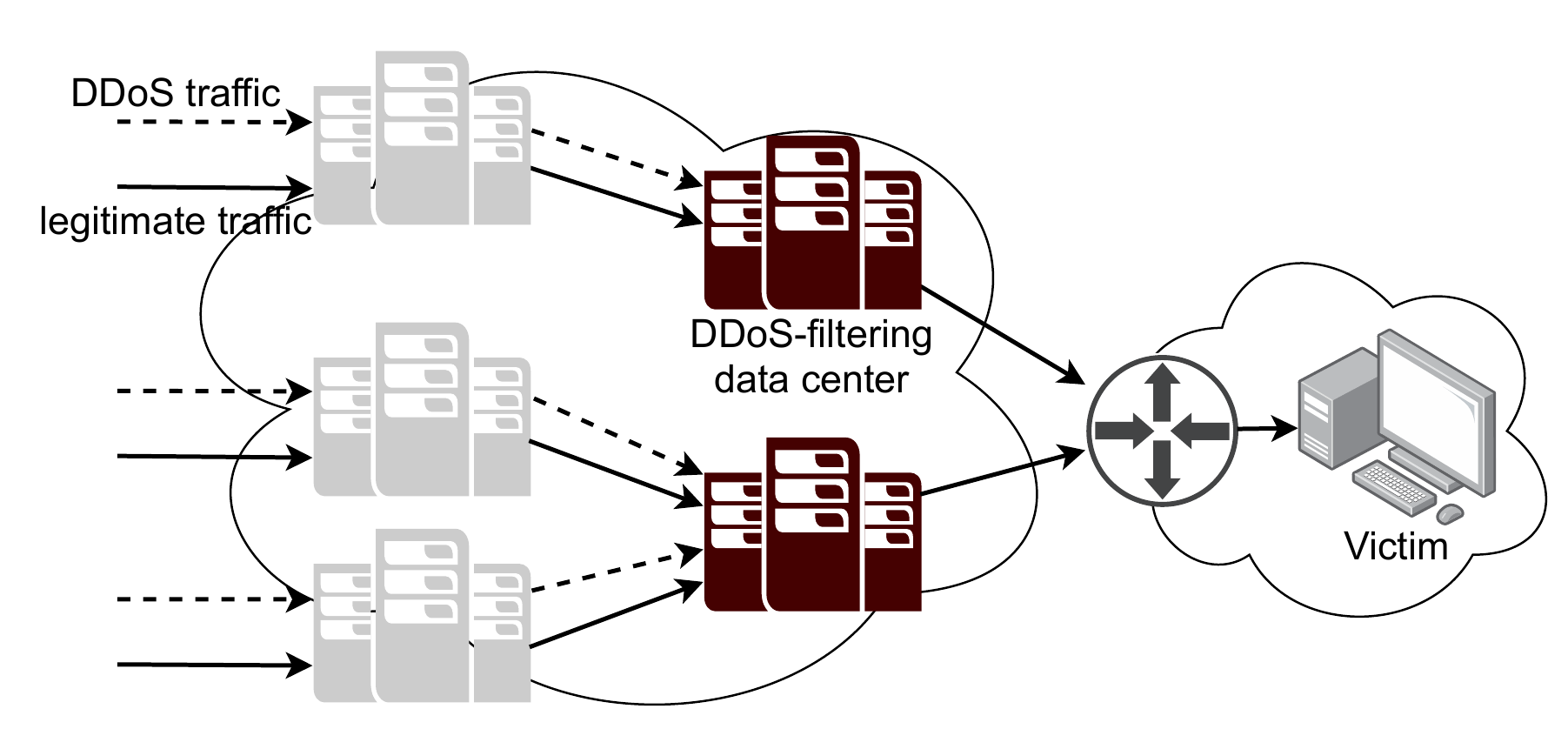}\label{fig:scrubbing}
	  }
   \caption{
	   \textbf{Distributed filtering at different scales.} 
	}
	\label{fig:distributed-filtering-model}
\end{figure}

The distributed filtering model is suitable at multiple scales, 
including the Internet scale, an ISP scale, or a scrubbing service scale.
The DDoS traffic can be filtered on their paths toward the victim
at different autonomous systems (ASes) throughout the whole Internet (Figure~\ref{fig:filter_as}),
or at different routers or programmable switches inside an ISP (Figure~\ref{fig:filter_router}),
or different data centers inside a DDoS-scrubbing service (Figure~\ref{fig:scrubbing}).
In this paper we focus on Internet-scale filtering of DDoS traffic \emph{en route},
but in general our design is also applicable to distributed filtering of DDoS at an ISP or scrubbing center level.

\subsection{Subscriber}

Distributed filtering of DDoS traffic can be viewed as a service
collectively provided by nodes that filter DDoS traffic.
Its clients can be a DDoS victim itself, or a node that acts on behalf of the victim,
such as the firewall of an institutional network or the upstream ISP of the victim.
We also call any client of a distributed filtering service a \textbf{subscriber}.

A subscriber's job is to determine what traffic needs to be filtered 
and select filtering nodes to filter such traffic. 
For the Internet-scale filtering of DDoS traffic,
the filtering nodes a subscriber selects will be ASes on the Internet, 
not routers or mitigation appliances in those ASes,
as it is up to an AS to determine its internal locations of filtering DDoS traffic.

DDoS defense is usually composed of three complementary processes 
that also map to three lines of research in parallel over the last two decades: 
DDoS detection and DDoS traffic classification that detects DDoS attack and classifies DDoS traffic and legitimate traffic;
DDoS path discovery that discovers the paths of DDoS traffic 
(or more generally their footprints that also include the source IP addresses and bandwidth consumption etc.);
and DDoS mitigation that filters, throttles, or redirects DDoS traffic.
This paper is focused on DDoS mitigation via adaptive distributed filtering.
Below we explain how our design leverages DDoS detection and classification and DDoS path discovery,
including its certain assumptions about them.

\subsection{DDoS Detection and Classification and DDoS Path Discovery}
\label{sec:model.ddos_traffic}

We assume a subscriber can detect DDoS attacks and classify ``flows'' to be DDoS flows or legitimate flows 
with a usable accuracy that commodity detection software can easily provide,
and it continuously does so.
A flow can be defined using any reasonable combination of the following attributes that can be derived from packet header fields:
\bi
\item Source: which can be an IP prefix, an IP address, an IP address plus a port number, or a wildcard;
\item Protocol: which can be TCP, UDP, IPSec (AH or ESP), ICMP, or a wildcard;
\item TCP flags: which can be SYN, SYN-ACK, ACK, FIN, or RST that indicates the status of a TCP segment, or a wildcard; and
\item Destination: which is the victim's IP prefix (if an entire subnet is under DDoS attack), 
	an IP address, or an IP address plus a port number (if a specific service is under DDoS attack).
\ei
The following are examples labeling a DDoS traffic flow:
\be
\item A general direct, protocol-agnostic, volumetric attack. Source = IP 1: Port 1, IP 1: Port 2, IP 3: Port 3, ..., IP 100, IP 101, ..., 
					IP Prefix 1, IP Prefix 2, IP Prefix 3, ...
\item An NTP amplification attack. Source = *: 123, Protocol = UDP.
\item A DNS flood. Source = *: 53, Protocol = UDP.
\item A TCP SYN flood. Source = *, Protocol = TCP, TCP Flag = SYN.
\item An ICMP flood. Source = *, Protocol = ICMP, ICMP Type = Echo request.
\ee

A subscriber can plug in a third-party DDoS detection software such as FastNetMon~\cite{fastnetmon}.
In doing so, note the software does not need to detect and classify DDoS traffic with 100\% recall or precision,
as DDoS mitigation does not need to filter every DDoS flow.
Such a software can be based on traffic measurement results and network capacities 
to set up threshold values for every ``flow'' as defined above 
on various network traffic statistics,
such as total byte or packet numbers, inbound to outbound traffic ratios, number of SYN packets, etc.
During the traffic monitoring, if a threshold for a ``flow'' is breached, 
the detection software can raise a DDoS alarm.
Furthermore, it can classify the flow, or its top subflows based on their risk level, as a DDoS flow.
For example, if traffic volume from an IP prefix is above the threshold,
it can label all traffic from the prefix as a DDoS flow,
or label the most volumetric connections from the prefix as DDoS flows,
so long as the total volume of the rest of the connections from the prefix is below the threshold.

We also assume a subscriber can track the DDoS traffic,
such as knowing the paths of a DDoS flow before they reach the victim,
so it can select the most suitable filtering nodes along the paths to filter the DDoS traffic.
Example solutions include those using marking techniques~\cite{savage2000practical,yaar2005fit,argyraki2005active} and
those based on logging~\cite{snoeren2001hash,li2004large,shi22pathfinder}.

\subsection{Adaptive Filtering}
\label{sec:model.adaptive}

Once a subscriber detects DDoS ``flows'', 
it then can request filtering nodes on the path(s) of these flows to filter them.
Certain types of DDoS traffic are straightforward to filter,
including those flows defined by the Protocol, TCP flags, and/or Destination attributes.
However, flows that are defined by different source attributes, with or without other attributes,
are challenging to handle.
Such flows correspond to three different filtering granularities:
\bi
\item \textbf{IP-prefix-based filtering} that discards all traffic from an IP prefix.
\item \textbf{IP-address-based filtering} that discards all traffic from an IP address. 
\item \textbf{IP-and-port-based filtering} that discards all traffic from an IP address with a given source port number.
\ei

All three filtering granularities have their advantages and disadvantages.
IP-prefix-based filtering results in the least number of DDoS ``flows'' to filter, 
\ie the least number of filtering rules as every DDoS flow maps to a filtering rule.
It thus in turn leads to least networking overhead in shipping rules to wherever they need to be deployed,
least memory overhead in storing rules, and 
least management overhead in deploying, monitoring and revoking rules.
It could also lead to faster deployment of all the rules and, with less rules to search, 
better performance in matching every DDoS packet to a rule and taking actions.
However, 
IP-prefix-based filtering may lead to collateral damage, sometimes perhaps even severe,
when traffic from a legitimate IP in an IP prefix is filtered.
Nonetheless, certain amount of collateral damage may be still acceptable to a subscriber, 
especially when it is under a severe DDoS attack.
IP-address-based filtering will cause less collateral damage,
but it can still happen if there is also legitimate traffic from the same IP address of a DDoS bot,
or worse, if a DDoS bot spoofs the IP address of a benign host
who happens to be also sending traffic to the victim.
IP-and-port-based filtering has the least possibility of collateral damage.
It also makes IP spoofing hard to succeed,
unless a DDoS bot can spoof both the IP address and the source port number of an active legitimate flow with the victim,
the chance of which is extremely slim.
However, IP-and-port-based filtering usually does not scale.

We thus introduce adaptive filtering to seek the best tradeoff among all the competing factors.
In particular, a subscriber can enforce filtering at different granularities.
A simple adaptive filtering strategy could be as follows.
For an IP prefix that originates DDoS traffic,
if the volume of legitimate traffic from the prefix is low,
assuming the victim is under a severe DDoS attack and can afford losing some legitimate traffic,
a rule that filters the entire IP prefix is probably applicable.
Otherwise, we can look at every sub-prefix of the prefix.
We can generate a rule for every sub-prefix that primarily originates DDoS traffic,
skip every sub-prefix that primarily originates legitimate traffic, and
apply the same filtering strategy here recursively on every sub-prefix that originates both DDoS and legitimate traffic.
If in this recursive process a sub-prefix becomes an IP address that originates both DDoS and legitimate traffic,
we can check which ports of the IP address originates DDoS traffic,
and only filter traffic from those ports of the IP address.

With adaptive filtering, a subscriber is no longer limited to a single granularity of filtering DDoS traffic,
such as AITF that filters DDoS traffic using individual IP addresses~\cite{Argyraki2005}.
Instead, a subscriber is able to elect to use different filtering granularities as needed.

The aforementioned simple adaptive filtering strategy, however, leaves many key questions unanswered.
A major challenge is that there can be thousands of DDoS flows from different IP prefixes,
millions of DDoS flows from different IP addresses,
and potentially even more from different IP address and port number combinations.
It will be prohibitively expensive to monitor traffic for every granularity and then determine filtering rules accordingly.
Moreover, it does not consider a subscriber's preferences in traffic filtering,
such as its objectives and constraints in terms of DDoS traffic coverage, collateral damage, and the number of rules to generate and deploy.
Also, for a flow from an IP prefix or IP address, 
it does not take advantage of the paths of the flow that a subscriber can learn (Section~\ref{sec:model.ddos_traffic}).
For example, if traffic from an IP address consists of DDoS traffic from one path and legitimate traffic from another distinct path,
we can employ IP-address-based filtering at a node that is on the former path but not on the latter path,
without resorting to the more specific but less scalable IP-and-port-based filtering. 
Finally, the simple adaptive filtering strategy is a top-down approach, moving from IP prefixes to sub-prefixes to IP addresses and then to ports.
However, DDoS detection and classification solutions usually classifies traffic flows into a fine granularity such that
every flow is either DDoS traffic or legitimate traffic (rather than a mixture of both) (Section~\ref{sec:model.ddos_traffic}).
To run adaptive filtering on top of DDoS classification,
it is more natural for adaptive filtering to be bottom-up instead. 
We address all these questions in the next section (Section~\ref{sec:system}).
\section{System Design} 
\label{sec:system}

Ideally, a subscriber wants to generate rules that are optimal for three objectives,
including a full coverage of DDoS traffic,
no collateral damage from dropping legitimate traffic, and
only using a small number of rules.
In practice, however, a subscriber must compromise one or two objectives in order to optimize for another objective,
and each subscriber may have a different set of prioritized objectives. 
In our design, we allow a subscriber to optimize for one objective,
but it must also meet the constraints for other objectives.

\subsection{Problem Formulation}
\label{sec:system.problem}

We now formulate the problem of rule generation. 
For a given rule $r$, we define $d$($r$, $T$) and $l$($r$, $T$)
to be respectively the DDoS traffic and legitimate traffic that rule $r$ filters from the traffic set $T$, respectively.
As such, if we have a set of rules $R$=$\{r_i|i$=1, $\dots$, $n\}$, where $r_i$ is a rule,
we have $d(R,T)$=$\sum_{i=1}^n d(r_i, T)$ and $l(R,T)$=$\sum_{i=1}^n l(r_i, T)$
to respectively represent the DDoS traffic coverage and collateral damage of the rule set $R$ over traffic $T$.
Assuming the subscriber's constraints for
the minimal amount of DDoS traffic that must be filtered is $D$, 
the maximal amount of legitimate traffic that could be filtered is $L$, and 
the maximal number of rules that is allowed to generate and deploy, which we also call \textbf{rule budget}, is $M$, 
we define three distinct single-objective rule-generation problems as follows:

\bb
\item \textbf{Rule-generation Problem 1:} 
In case the subscriber is most concerned about filtering as much DDoS traffic as possible,
for traffic $T$, output a set of rules $R$=\{$r_i|i=1,...,n\}$ that
maximizes $d(R, T)$, whereas $l(R, T)$$\leq$$L$ and $|R|$$\leq$$M$. 
Example scenario 1: the victim is overwhelmed by a severe DDoS attack and eager to have as much DDoS traffic as possible filtered.

\item \textbf{Rule-generation Problem 2:}
In case the subscriber is most concerned about 
avoiding collateral damage due to the filtering of legitimate traffic,
for traffic $T$, output a set of rules $R$=\{$r_i|i=1,...,n\}$ that
minimizes $l(R, T)$, whereas $d(R, T)$$\geq$$D$ and $|R|$$\leq$$M$;
Example scenario 2: the DDoS attack is not that severe, and the victim does not wish legitimate traffic to be filtered by mistake.

\item \textbf{Rule-generation Problem 3:}
In case the subscriber is most concerned about minimizing the number of generated rules,
for traffic $T$, output a set of rules $R$=\{$r_i|i=1,...,n\}$ that
minimizes $|R|$, whereas $l(R, T)$$\leq$$L$ and $d(R, T)$$\geq$$D$.
Example scenario 3: deploying filtering rules costs a certain amount of money,
and the subscriber may have a limited budget to defend against an attack.

\eb

The subscriber then can choose which problem to solve,
depending on which metric to optimize and which metrics to impose constraints.

\subsection{$F$-tree}
\label{sec:system.f-tree}

We now describe a data structure called $F$-tree, 
which we will use to generate DDoS-filtering rules as described in Section~\ref{sec:system.algo}.
An $F$-tree is a tree in which every node records a traffic source\consider{that may need to be filtered} 
and every parent node records an aggregated source that aggregates all the sources represented by its child nodes.
Specifically, every node in an $F$-tree records the following information of a source:
\bi
\item $S$: The source of traffic\consider{that may need to be filtered}. It can be an IP source address and a port number, an IP address, or an IP prefix.
	We call all packets from $S$ toward the victim a ``flow'' from $S$.
	Note the source is not necessarily a single end point on the Internet.
	Even if it is a single IP address, because of IP spoofing, there may be more than one path. 
\item $F$: A set of candidate filtering nodes on the path(s) of the flow that may be used to filter packets from the flow. 
\item $d$: The amount of DDoS traffic from the flow in terms of number of bytes, packets, or TCP or UDP connections, that can be filtered by $F$.
	This is also the DDoS coverage when using a node from $F$ to filter traffic from $S$.
\item $l$: The amount of legitimate traffic from the flow that can be filtered by $F$.
	This is also the collateral damage when using a node from $F$ to filter traffic from $S$.
\ei

\begin{figure}[ht]
	\centering
	\subfigure[Parent node ($N_1$) with two children ($c_1$, $c_2$) via union aggregation]{%
	\includegraphics[width=0.26\textwidth]{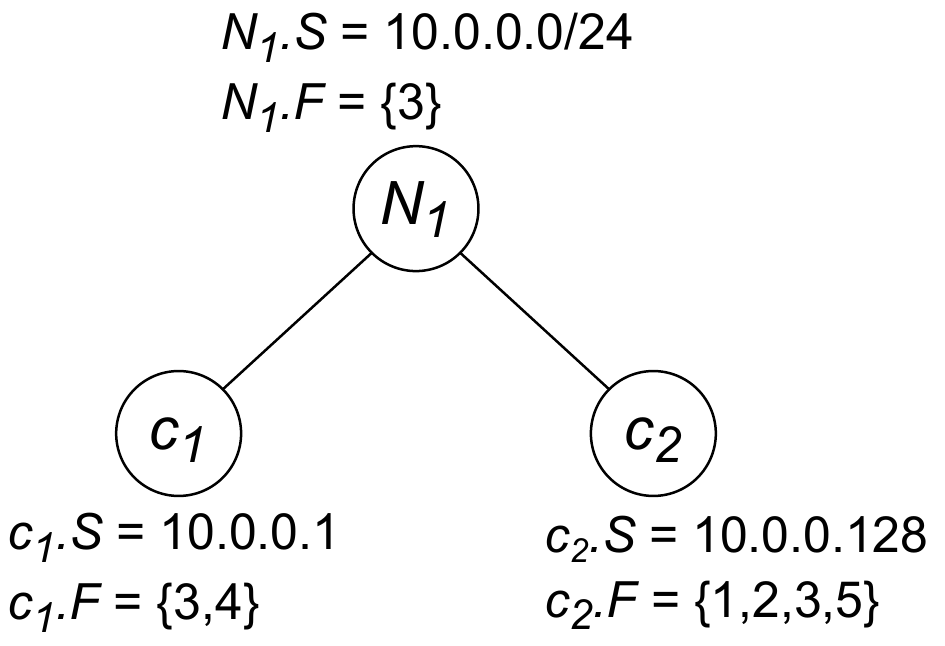}
		\label{fig:F-tree-example_a}}
	\subfigure[Parent node ($N_2$) with two children ($c_3$, $c_4$) via difference aggregation]{%
	\includegraphics[width=0.3\textwidth]{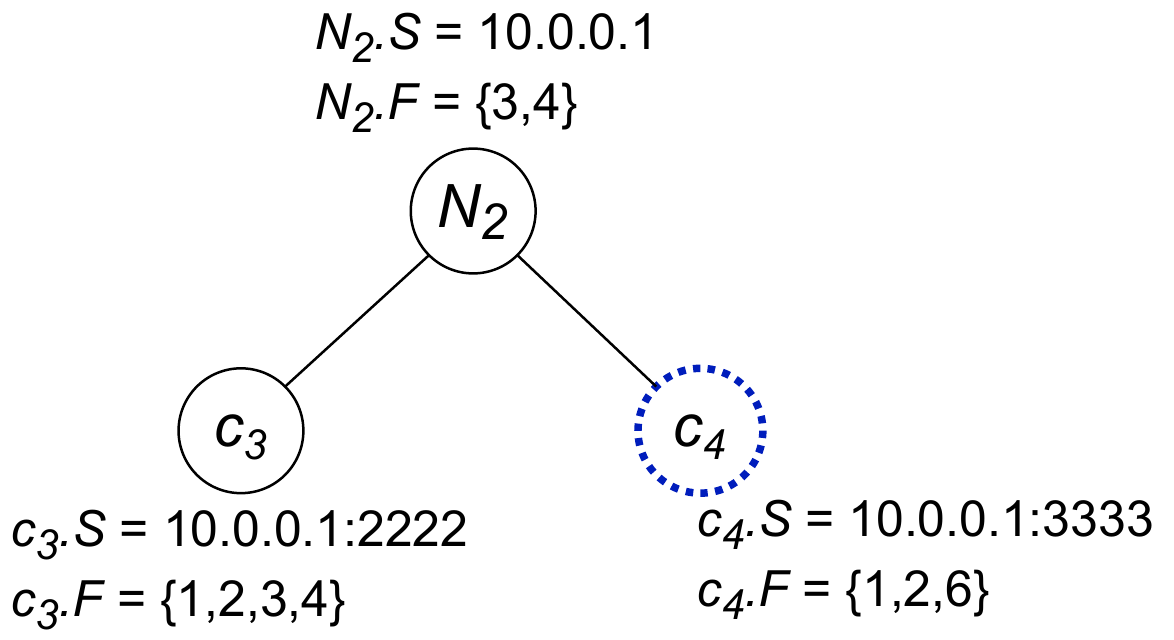}
		\label{fig:F-tree-example_b}}
	\subfigure[The distributed filtering topology behind the two $F$-tree examples above]{%
	\includegraphics[width=0.27\textwidth]{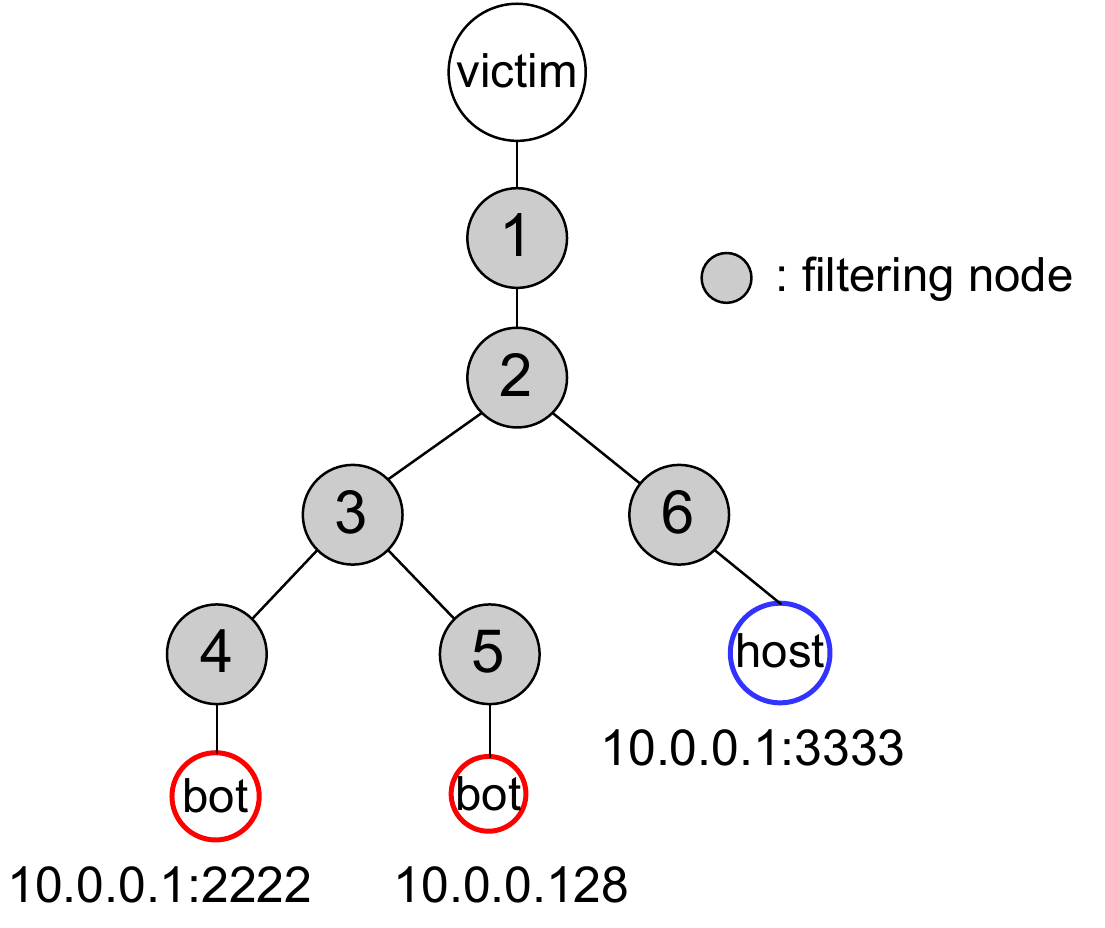}
		\label{fig:F-tree-example_c}}
	\caption{\textbf{F-tree for DDoS-filtering rule generation.}}
	\label{fig:F-tree-example}
\end{figure}

Basically, every node $N$ on an $F$-tree can map to a rule that, 
if deployed on one of the nodes in $N.F$, can drop traffic whose source addresses matches $N.S$,
with a DDoS coverage of $N.d$ and collateral damage of $N.l$.
Figures~\ref{fig:F-tree-example_a} and \ref{fig:F-tree-example_b} show two toy $F$-tree examples. 

Every node $N$ with a set of child nodes $c_1, ..., c_n$ (in a binary tree $n$ is 1 or 2)
derives its four values from those of its children through aggregation.
First, the source value of $N$ is the aggregation of the source values of all its child nodes.
Specifically, $N.S$=\textit{prefix}({$c_1.S, \dots, c_n.S$}), where \textit{prefix()} is a function 
to extract the longest\consider{ (\ie the most specific)} common prefix from input prefixes. 
For example, in Figure~\ref{fig:F-tree-example_a},
if node $N_1$ has two children $c_1$ and $c_2$, $c_1.S=10.0.0.1$ and $c_2.S=10.0.0.128$, then $N_1.S=10.0.0.0/24$.
Or for another example, in Figure~\ref{fig:F-tree-example_b}, 
if node $N_2$ has two children $c_3$ and $c_4$, $c_3.S=10.0.0.1:2222$, $c_4.S=10.0.0.1:3333$, where 2222 and 3333 are source port numbers,
then $N_2.S=10.0.0.1$.

There are two types of aggregation: \textbf{union aggregation} or \textbf{difference aggregation}.
Both can only happen if they do not lead to an empty $N.F$.
A union aggregation is to derive information for filtering all the flows represented by child nodes.
It is as follows:
\bi
\item $N.F=\cap_{i=1}^n (c_i.F)$;
\item $N.d=\sum_{i=1}^n (c_i.d)$; and  
\item $N.l=\sum_{i=1}^n (c_i.l)$.
\ei
Assume node $N_1$ above is derived via a union aggregation. 
If $c_1.F=\{3,4\}$ and $c_2.F=\{1,2,3,5\}$, 
then $N_1.F=\{3\}$, $N_1.d$ = $c_1.d+c_2.d$, and $N_1.l$ = $c_1.l+c_2.l$.

A difference aggregation is to derive information for filtering only certain flows represented by child nodes
and avoid filtering certain flows represented by child nodes.
Assume among child nodes $c_1, ..., c_n$, we want to filter flows from $c_1, ..., c_k$ but not flows from $c_{k+1}, ..., c_n$,
a difference aggregation is as follows:
\bi
\item $N.F=\cap_{i=1}^k (c_i.F)$ - $\cup_{i=k+1}^n (c_i.F)$;
\item $N.d=\sum_{i=1}^k (c_i.d)$; and  
\item $N.l=\sum_{i=1}^k (c_i.l)$.
\ei
Assume node $N_2$ above is derived via a difference aggregation. 
If $c_3.F=\{1,2,3,4\}$, $c_4.F=\{1,2,6\}$, and $N_2$ wants to filter traffic from $c_3$ but not $c_4$,
then $N_2.F=\{3, 4\}$, $N_2.d$ = $c_3.d$, and $N_2.l$ = $c_3.l$.

Finally, if we combine our above examples for nodes $N_1$ and $N_2$ and map $N_2$ to $c_1$ (they have the same values),
we can obtain a bigger $F$-tree.
Figure~\ref{fig:F-tree-example_c} shows the underlying topology.
We can see if we want to filter DDoS traffic from $10.0.0.1:2222$ ($c_3$) and $10.0.0.128$ ($c_2$)
without a collateral damage on traffic from $10.0.0.1:3333$ ($c_4$),
we will obtain a rule represented by $N_1$, \ie filter traffic from $10.0.0.0/24$,
to be deployed in one of nodes in $N1.F$, \ie node 3.

\subsection{Rule Generation Algorithm}
\label{sec:system.algo}

We now describe how a subscriber generates rules using an F-tree.
First, as a DDoS victim continuously receives traffic,
the subscriber acting on behalf of the victim can classify/label incoming traffic flows to be DDoS flows or legitimate flows,
and also know the nodes on the path(s) of the flows that can filter traffic (Section~\ref{sec:model.ddos_traffic}). 
With such information for every incoming flow, the subscriber can accordingly initialize all the leaf nodes in the F-tree. 
For all labeled traffic from the same source,
the subscriber casts them into a leaf node, say $N$, on the F-tree,
where $N.S$ is the source, 
$N.F$ are all the filtering nodes on the path of traffic from $N.S$,
and $N.d$ and $N.l$ are the amount of DDoS and legitimate traffic from $N.S$, respectively.
($N.d$ and $N.l$ are respectively zero for legitimate and DDoS flows.)
It then runs a loop process which recursively aggregates leaf nodes to generate parent nodes,
following the aggregation procedure in Section~\ref{sec:system.f-tree}.
The key at every iteration of the loop is to determine which nodes to aggregate 
based on the rule-generation problem in place, as follows.

For the rule-generation problem 1 that maximizes the DDoS coverage,
in each iteration, the algorithm first finds leaf nodes, if aggregated, 
that will bring the highest increase of the DDoS coverage without violating the collateral damage constraint.
It then derives their parent node as described in Section~\ref{sec:system.f-tree},
prunes the leaf nodes,
and makes the parent node a new leaf node.
The loop process continues until no such aggregation can be done.
The subscriber then maps the top up to $M$ leaf nodes with the highest $d$-values to the rules to use.

For the rule-generation problem 2 that minimizes the collateral damage,
in each iteration, the algorithm first finds leaf nodes, if aggregated, 
that will introduce the least collateral damage. 
It then derives their parent node, 
prunes the leaf nodes,
and makes the parent node a new leaf node.
The loop process continues until in the current F-tree there are $M$ or fewer leaf nodes 
whose sum of $d$ values are at least $D$.
It then maps these $M$ or fewer leaf nodes to the rules to use.

Finally, for the rule-generation problem 3 that minimizes the number of rules,
in each iteration, the algorithm first finds the largest number of leaf nodes 
whose aggregation into a parent node will not violate the collateral damage constraint.
It then derives their parent node, 
prunes the leaf nodes,
and makes the parent node a new leaf node.
The loop process continues until no such aggregation can be done.
It then returns the least number of leaf nodes 
whose total collateral damage is less than $L$ and total DDoS coverage is at least $D$,
and maps these leaf nodes to the rules to use.

\subsection{Rule Placement}
\label{sec:system.placement}

Once rules are generated, the subscriber can inspect all the rules and deploy them.
For every rule, it can look at the $F$-tree node that corresponds to the rule, say $N$, 
and choose one of the filtering nodes in $N.F$ to place the rule.
It then can contact the node for rule placement;
if the node is unavailable, the subscriber can choose another node in $N.F$ for rule placement. 
If no node in $N.F$ can place the rule, this rule cannot be placed.
The subscriber can first try to deploy rules that only have a single possible deployment location ($|N.F| = 1$),
and then those with two locations, and so on.

\subsection{Rule Generation Complexity Analysis}
\label{sec:system.complexity}

The rule-generation algorithm targets three rule generation problems with a similar structure. 
Its main loop begins with an initial set of leaf nodes, and it performs an aggregation at each iteration
until a stopping condition is satisfied.
Assuming that the number of leaf nodes is initially $n$,
the maximum number of aggregations that can be performed is then $n$-$1$, 
and thus the main loop may run $n$-$1$ times in the worst case.
Within this main loop, it is necessary to sort the leaf nodes,
which dominates other terms within the loop.
With a sorting complexity of $O(nlog(n))$ in the first iteration to sort all $n$ leaf nodes
but only $O(n)$ in later iterations to insert a newly made leaf node into an already ordered list of other leaf nodes,
we have a \textit{worst-case} complexity of $O(n^2)$.

\section{Implementation}
\label{sec:implementation}

\subsection{Adaptive Distributed Filtering Software Suite} 
\label{sec:implementation.software}

We have developed an adaptive distributed filtering software suite 
composed of a set of independent applications,
including the subscriber application and the filtering-node application.
The subscriber application takes classified/labeled traffic as input and 
includes modules on rule generation and rule placement.
The filtering-node application can interact with a wide range of filtering capabilities,
including BGP FlowSpec, Cisco ACL, and all major SDN controller software
(\eg OpenDaylight\cite{opendaylight2018}, ONOS\cite{berde2014onos}, and Ryu\cite{ryu2017}).

\subsection{Adaptive Distributed Filtering Protocol} 
\label{sec:implementation.protocol}

We developed a protocol to define the messages between any subscriber and any filtering node.
The most important message type is \textbf{rule submission}.
It includes a version number, a message type, the rule ID,
and the rule itself that is defined by the four fields
(source, protocol, TCP flags, and destination) 
described in Section~\ref{sec:model.ddos_traffic}.
Further, it includes a starting time field regarding when the rule should start taking effect
and an ending time field indicating when the rule should expire. 
Also, we define a type of message called \textbf{rule submission acknowledgment},
which is sent in response to a rule submission message.
It also contains a version number, a message type, a rule ID that is the ID of the rule being acknowledged.
Moreover, it includes an error code that indicates either the rule is installed successfully (with error code being zero),
or the reason the rule is not installed successfully,
including verification error, timing error, out of rule space, internal error, or other.

\subsection{Security and Privacy Considerations}
\label{sec:implementation.security}

To ensure our system is not misused or abused, we tackle the following security and privacy issues: 

\emph{Privacy:}
The most essential information sharing in the system is that 
when a subscriber runs the rule generation process,
it may learn newly detected DDoS flows and their paths from the DDoS detection, classification and tracking components.
It may also learn who the filtering nodes are on each path. 
However, as these paths are the same paths meant to be announced and propagated for packet routing on the Internet,
there is thus no privacy concern here. 

\emph{Authentication:} 
Every party in the system (filtering nodes, subscriber, or DDoS victim) 
must have a signed, verifiable certificate 
so that other parties can verify its identity, IP address(es), public key, and other metadata.
If it does not already have a certificate signed by an existing public-key infrastructure (PKI) that our system recognizes,
it can obtain a certificate from an internal PKI of our system through a registration process.

\emph{Traffic Ownership:}
When a subscriber requests a filtering node to deploy a rule,
the rule must only filter traffic to the victim that the subscriber represents.
To ensure this, our system mandates that the subscriber have a \textbf{traffic control authorization (TCA)} ticket
issued and signed by the victim to prove that the subscriber is allowed to issue rules against traffic to the victim.
The subscriber must also sign the rule.
This design also allows our system to protect any link of an ISP from the Crossfire DDoS~\cite{kang13crossfire}:
the ISP can contact each downstream network beforehand for a TCA ticket to become its subscriber
and then generate and deploy rules to filter the Crossfire traffic toward each downstream network.

\emph{Message Protection:}
Communication within our system must achieve confidentiality, integrity, and authentication.
Each communication channel will leverage the certificate obtained from the PKI to open an HTTPS connection.
Since HTTPS builds upon the Transport Layer Security (TLS) protocol,
the communication channels will achieve the desired message protection goals. 
\section{Evaluation}
\label{sec:evaluation}

\subsection{Overview}
\label{sec:evaluation.setup}

We built a simulation platform consisting of the actual implementation of our system 
and a simulation of the Internet data plane.
We measured our system's ability at the Internet scale to defend a victim against real-world, large-scale DDoS attacks.
We replayed three real-world DDoS attacks of different sizes and attack dynamics:
RADB-DDoS~\cite{radb2016} with the DNS protocol and $\sim$16,000 DDoS sources, 
Booter1-DDoS~\cite{santannaBooters} with the DNS protocol and $\sim$4,500 DDoS sources, and 
CAIDA-DDoS~\cite{caida2007} with the ICMP protocol and $\sim$7,000 DDoS sources.

We first evaluated
rule generation in Section~\ref{sec:evaluation:rule-gen}, 
focusing on the efficacy
and tradeoffs of different rule generation objectives,
and rule deployment in Section~\ref{sec:evaluation:deployment}, 
focusing on the percentage of rules for which suitable locations are found (\ie success rate) and 
the distribution of rules across participating ASes. 
In Section~\ref{sec:evaluation:e2e}, we assessed our system's efficacy in mitigating DDoS in real time.
Lastly, in Section~\ref{sec:eval.deployment} we deployed and evaluated our system on the GENI testbed~\cite{GENI} and two real-world IXPs.

\subsection{Rule Generation}
\label{sec:evaluation:rule-gen}

We evaluated the rule generation algorithm and measured the resulting DDoS coverage, collateral damage, and number of generated rules
for each of the rule-generation problems described in Section~\ref{sec:system},
and compared the tradeoffs of differing rule generation strategies. 
We focused on rules based on source IP addresses of the traffic.
While rule generation is a continuous process running in real time and handles a batch of DDoS and legitimate flows each time,
we focused on one batch of traffic over a second composed of 1000 attack sources and 500 legitimate sources, all randomly generated. 
The size of the batch is not too big to cause a slow response with many DDoS flows,
but not too small either to result in too many batches. 

\begin{figure}[t]
 \centering
 \subfigure[Maximizing DDoS Coverage vs. L and M] {
   \includegraphics[width=0.32\textwidth]{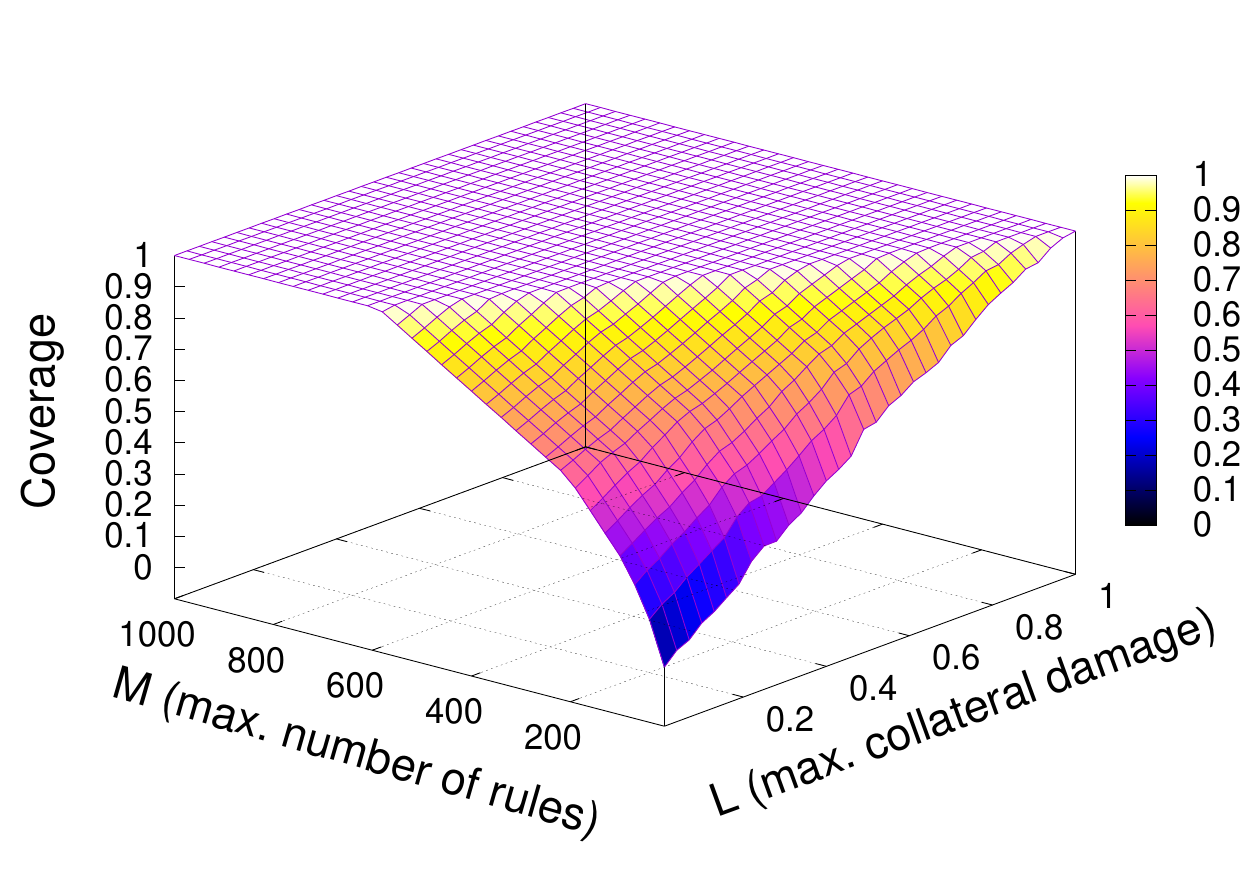}
   \label{fig:rule-gen-coverage}}
 \vspace{-1mm}
 \subfigure[Minimizing Collateral damage vs. D and M] {
   \includegraphics[width=0.32\textwidth]{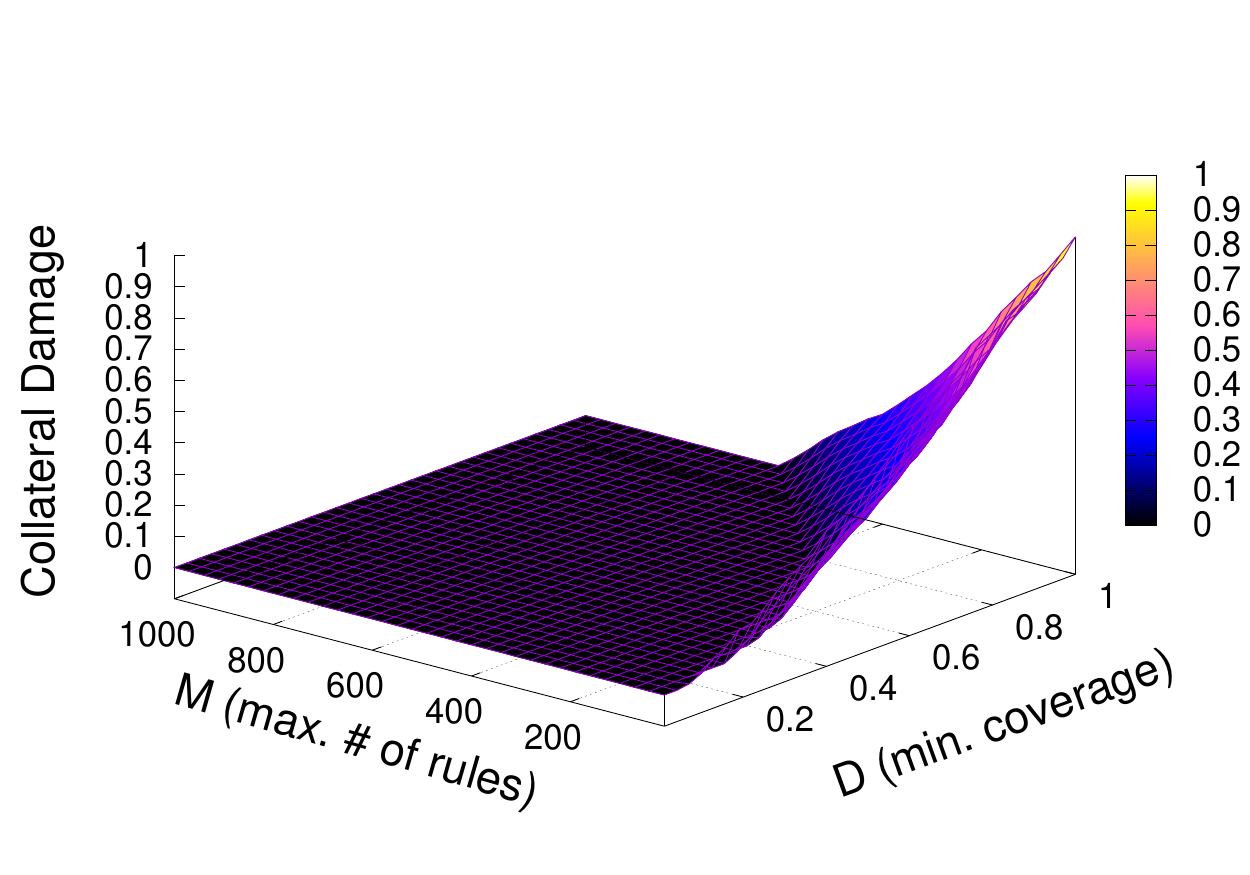}
   \label{fig:rule-gen-collateral}}
 \vspace{-1mm}
 \subfigure[Minimizing Number of Rules vs. D and L] {
   \includegraphics[width=0.32\textwidth]{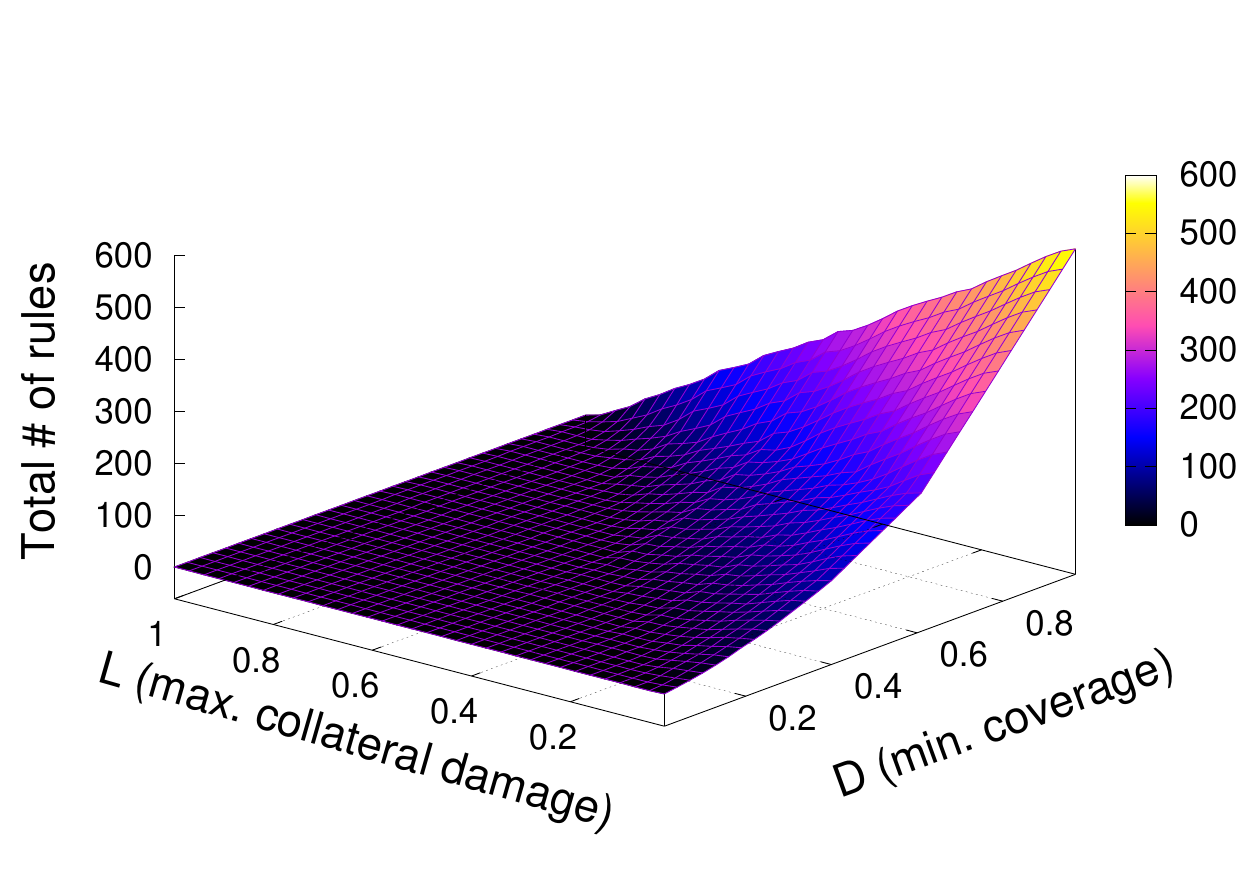}
   \label{fig:rule-gen-rule-numbers}}
\caption{\textbf{Rule generation with constraints} ($D$: minimum DDoS coverage; $L$: maximal collateral damage;  $M$: rule budget).}
\end{figure}

We found the algorithm achieves optimal results for all three rule-generation problems (1, 2 and 3).
We first examined our algorithm for rule-generation problem 1 described Section~\ref{sec:system}.
Here, the goal is to maximize the DDoS coverage, while satisfying constraints
on the maximum number of rules $M$ and the maximum amount of acceptable collateral damage $L$.
We vary the values for $L$ and $M$ and examine the DDoS coverage.
As shown in Figure~\ref{fig:rule-gen-coverage},
100\% DDoS coverage is achieved easily, except when $L$ and $M$ are both low.
In these cases, however, the DDoS coverage is still maximized subject to the stringent constraints on $L$ and $M$.

We then examined our algorithm for rule-generation problem 2 described in Section~\ref{sec:system}.
The goal of this algorithm is to minimize the collateral damage, 
subject to constraints on the minimum DDoS coverage and maximum number of rules.
Figure~\ref{fig:rule-gen-collateral} shows that
the collateral damage
vary as expected according to the values of the minimal DDoS coverage $D$ and the maximum number of rules $M$.
In particular, collateral damage is indeed minimized, and is zero in most cases.
When $D$ is high and $M$ is low, some collateral damage is incurred,
since the only way to cover a large percentage of unwanted flows
with a relatively small number of rules is to allow some collateral damage to occur.

Finally, we examined our algorithm for rule-generation problem 3 described in Section~\ref{sec:system}.
The goal of this algorithm is to minimize the number of rules, 
while satisfying the constraints on the minimum DDoS coverage and maximum acceptable collateral damage.
Figure~\ref{fig:rule-gen-rule-numbers} shows the results.
We can see that in most cases only one or a small number of rules are generated,
except when the minimum DDoS coverage ($D$) is high and the maximum collateral damage ($L$) is low.

\subsection{Rule Deployment}
\label{sec:evaluation:deployment}

Continuing with rules generated in Section~\ref{sec:evaluation:rule-gen}, 
we evaluated the distributed deployment of DDoS-filtering rules against 
a number of distinct, Internet-scale distributed filtering profiles. 
Each profile represents different rates of ASes on the Internet that participate distributed filtering of DDoS traffic,
as shown in Table~\ref{tab:DF-profiles}.
The total number of ASes in tiers 1, 2, and 3 is 89, 8442, and 47052, respectively.
Full-participation profile is clearly unrealistic, 
but we use this profile as a baseline.
The ``victim only'' profile serves as another baseline,
in which the victim's ISP is the only AS that filters DDoS traffic
and all rules must be deployed there.

\begin{table}[h]
  \centering
  \caption{\textbf{Distributed filtering profiles for rule deployment experiments.}
  (These numbers are the same as the real Internet.)
  }
  \begin{tabular}{l|rrrr}
    \toprule
    Name            & Tier 1 & Tier 2 & Tier 3 & Total \# \\
    \midrule
    Full-participation & 100\%  & 100\%  & 100\%  & 55583 \\ 
    Tier-1-only	    & 100\%  & 0\%    & 0\%    & 89 \\ 
    Top-centered    & 100\%  & 50\%   & 0\%    & 4310 \\ 
    Middle-centered & 0\%    & 80\%   & 20\%   & 16163 \\ 
    Bottom-centered & 0\%    & 20\%   & 80\%   & 39330 \\ 
    Victim-only     & 0\%    & 0\%    & 0\%    & 1 \\
	\bottomrule
  \end{tabular}
  \label{tab:DF-profiles}
\end{table}

\begin{figure}[t]
  \centering
  \includegraphics[width=0.9\columnwidth]{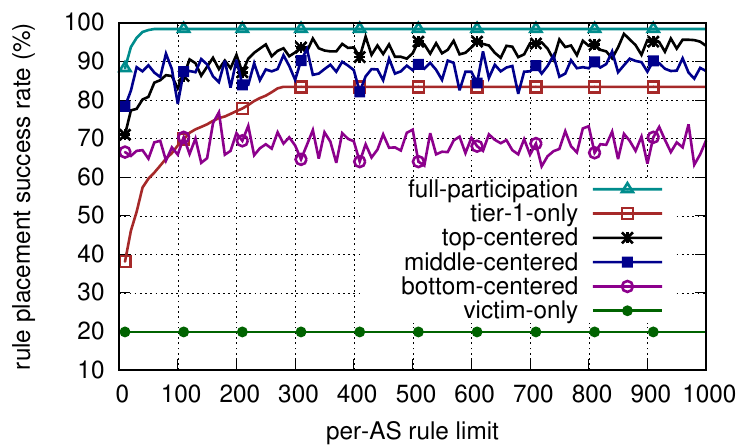}
  \caption{\textbf{Rule deployment success rates.}}
  \label{fig:rule-placement-overall}
\end{figure}

We first evaluated the rule deployment success rate, \ie the percentage of rules for which suitable locations are found.
With rules from Section~\ref{sec:evaluation:rule-gen} as input,
Figure~\ref{fig:rule-placement-overall} depicts the success rate under each profile.
The first and most obvious trend displayed is that the success rate 
for all profiles either remains stable or generally increases as we increase the per-AS rule limit from 1 to 1000.\consider{Clearly, 
more available rule space at deploying ASes results in fewer rule deployment failures due to exhausted space.}
Another trend
is the impact of a higher overall rate of ASes participating the distributed DDoS filtering.
Overall, the rule deployment success rate increases with a higher rate of participating ASes,
though increasing the participation rate for some AS tiers has different effects than for others.
As expected, the lowest success rate belongs to the victim-only profile, while
the highest rate is achieved by the full-participation profile.
The four profiles in between generally perform much better than the victim-only profile,
and slightly or moderately worse than the full-deployment profile, where the top-centered profile
is the only profile of these four to reach nearly 100\% success rate,
and generally performs better than the others.
The middle-centered profile is not far behind, however,
and actually reaches higher success rates than the top-centered profile
when the number of rules per AS is low.
The tier-1-only profile is the most sensitive to the per-AS rule limit, 
as with only 89 tier-1 ASes each AS faces pressure to deploy more rules than other profiles;
it thus has a lower success rate than other profiles (except for victim-only) when the per-AS rule limit is low,
but gradually improves as the limit gets higher.

\begin{figure}[h]
  \centering
  \includegraphics[width=0.9\columnwidth]{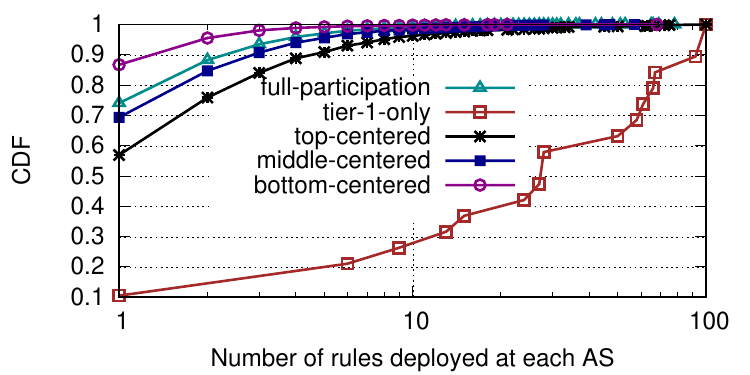}
  \caption{\textbf{Distribution of the number of rules at ASes participating distributed filtering ($\leq$100 rules per AS).}}
  \label{fig:rule-placement-CDF-100}
\end{figure}

Next, we examined how many rules are placed at each participating AS
under five different distributed filtering profiles 
while each AS can only deploy at most 100 rules
(Figure~\ref{fig:rule-placement-CDF-100}).
Across all profiles, except the tier-1-only profile,
approximately 60\% or more of ASes that participate in the defense must deploy only a single rule,
approximately 95\% or more of ASes deploy no more than 10 rules,
and thus a very small percentage of ASes deploy more than 10 rules.
For the tier-1-only profile, the rules are more spread out among all ASes,
but note this profile corresponds to the smallest actual number of ASes.
Overall, we can see that
the deployment of rules takes advantage of the fact that
a small number of ASes are in especially advantageous locations
and can contribute disproportionately to the defense.

\subsection{DDoS Mitigation}
\label{sec:evaluation:e2e}

\begin{figure}[th]
  \centering
  \subfigure[CAIDA-DDoS attack under rules for maximal coverage with varying rule budgets.] {
    \label{fig:e2e-sub1}
    \includegraphics[width=0.49\textwidth]{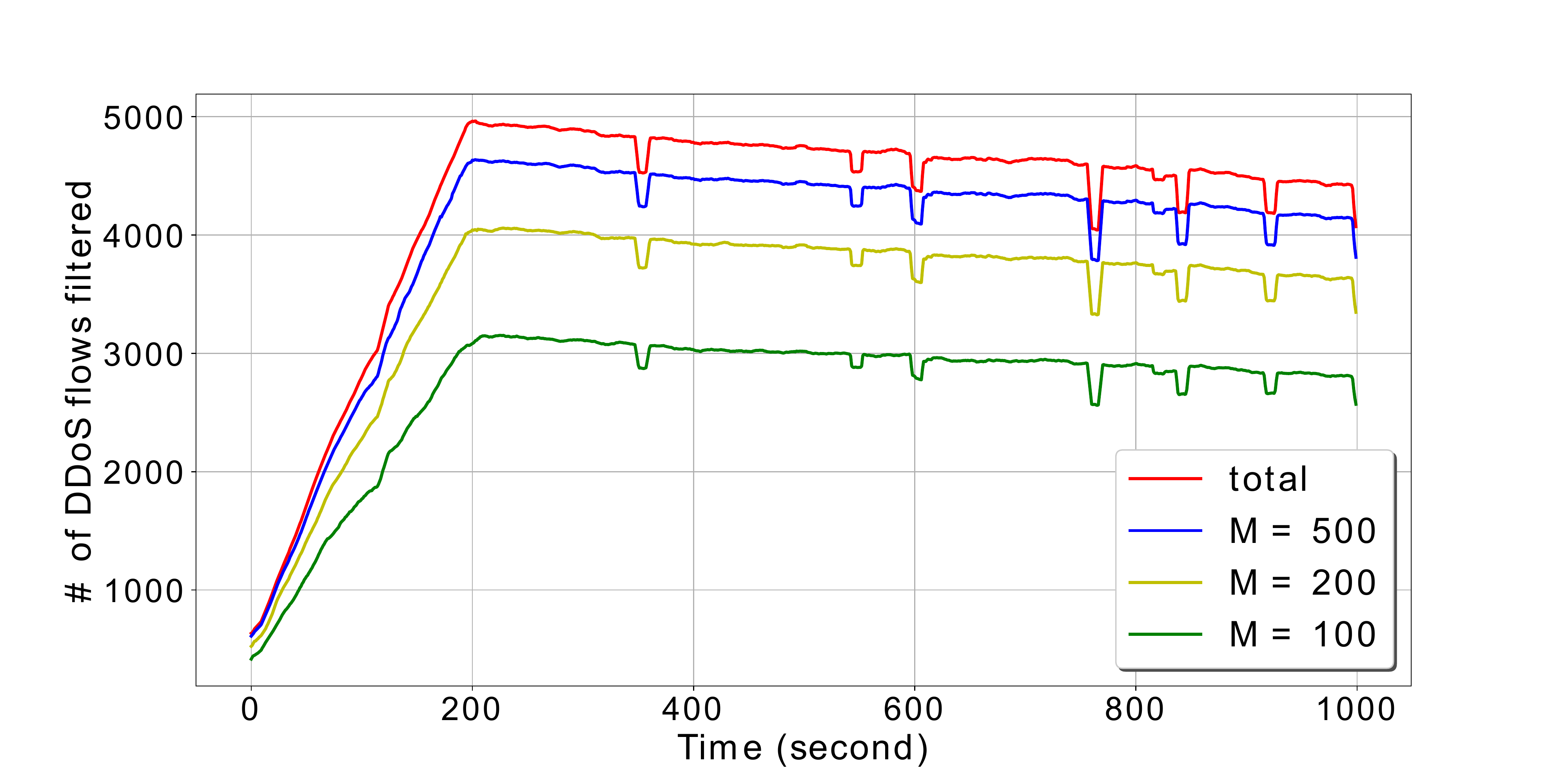}}
  \subfigure[RADB-DDoS attack under rules for minimal number of rules with varying coverage.] {
    \label{fig:e2e-sub2}
    \includegraphics[width=0.49\textwidth]{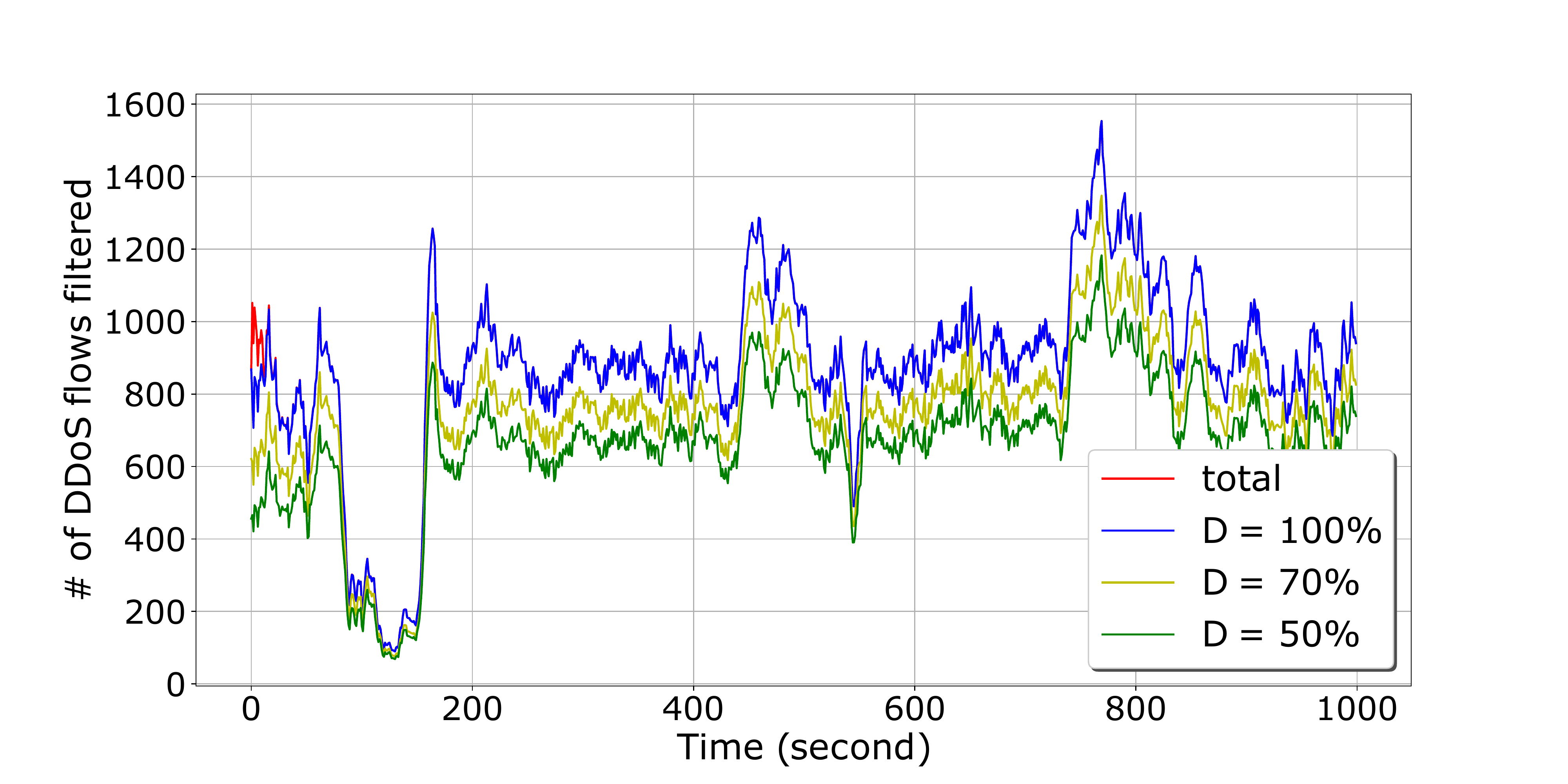}}
  \caption{\textbf{Time series of filtering of DDoS flows.} The ``total'' curve shows DDoS flows without filtering.}
  \label{fig:e2e-sub}
\end{figure}

We also evaluated the overall efficacy of adaptive distributed filtering
as we defend in real time against real-world DDoS attack traces
with continuous rule generation and placement.
Figure~\ref{fig:e2e-sub} shows two representative time series for our defense 
against two DDoS attacks with dissimilar dynamics (CAIDA-DDoS and RADB-DDoS).
For each attack, we show the number of DDoS flows filtered at each second during the attack
as well as the number of flows that arrive at the victim when no filtering is performed;
although not shown, no legitimate flows are ever filtered.

More specifically, Figure~\ref{fig:e2e-sub1} applies rules 
that are generated based on source addresses of the traffic
toward maximal DDoS coverage under \emph{zero} collateral damage requirement \emph{and} three different rule budgets
(100, 200, and 500, which represent roughly 1.5\%, 3\%, and 7\%, respectively, of the total approximately 7,000 DDoS sources).
Here, even with a tight budget of 100 source-based rules, which is only 1.5\% of DDoS sources,
60-70\% of DDoS flows will be filtered, 
and a higher value for the rule budget leads to more effective filtering. 

Figure~\ref{fig:e2e-sub2} instead applies rules 
that are generated toward minimum number of rules 
under \emph{zero} collateral damage requirement \emph{and} three different requirements on minimum DDoS coverage (100\%, 70\%, and 50\%).
The generation and placement of rules tracks very closely the spikes in the attack traffic,
demonstrating the overall accuracy of our rule generation algorithm.
In particular, with rules required to cover 100\% DDoS,
although initially not all DDoS flows are filtered,
it takes only about 13 seconds to begin filtering \emph{all} DDoS flows at every second afterwards.

\subsection{Pilot Studies}
\label{sec:eval.deployment}

We have deployed and tested a distributed DDoS filtering pilot system 
on the GENI (Global Environment for Network Innovations) testbed~\cite{GENI}.
Based on a recent Internet topology that consists of all Internet ASes, 
we chose a subgraph of 1 tier-1 AS, 18 tier-2 ASes, and 31-tier3 ASes 
where each of the total 50 ASes participates the filtering of DDoS traffic.
We also attached a local machine to one of the 50 ASes as a subscriber.
Each of these 50 ASes is supported with two virtual machines provided by GENI. 
The first virtual machine for each AS runs a Ryu controller as an SDN controller
and an Open vSwitch\cite{openvswitch} as an SDN switch that can deploy OpenFlow rules to filter traffic. 
The Open vSwitch is populated with a forwarding table by running the OSPF routing protocol~\cite{ospf}. 
The second virtual machine for each AS acts as an end-host in the AS
that can generate benign traffic toward a destination from different IP addresses of the AS.
More, in order to emulate large-scale DDoS attacks on the topology,
we installed a DDoS agent on each AS's second virtual machine.
It can receive commands about a variety of DDoS attacks from a bot master that we deployed on GENI and 
generate DDoS traffic toward a victim at a scheduled time from different IP addresses of the AS. 

Our system runs smoothly on this platform with good performance and low network overhead.
It also runs fast with rule generation at 105 milliseconds on average and
the network overhead is no more than 10 kilobytes each round for rule deployment.

Below we exemplify our system's effective filtering of DDoS traffic by launching an emulated
100-Gbps DDoS attack 
toward the subscriber from $\sim$1000 source addresses,
together with 40- to 60-Gbps legitimate traffic to the subscriber from $\sim$200 sources.
The subscriber will then generate rules based on its newly incoming DDoS traffic and
have these rules eventually converted to OpenFlow rules and 
deployed at selected Open vSwitches to filter the DDoS traffic.
 
\begin{figure}[th]
  \centering
  \subfigure[Traffic time series under rules for maximal DDoS coverage] {
    \label{fig:geni-1}
    \includegraphics[width=0.49\textwidth]{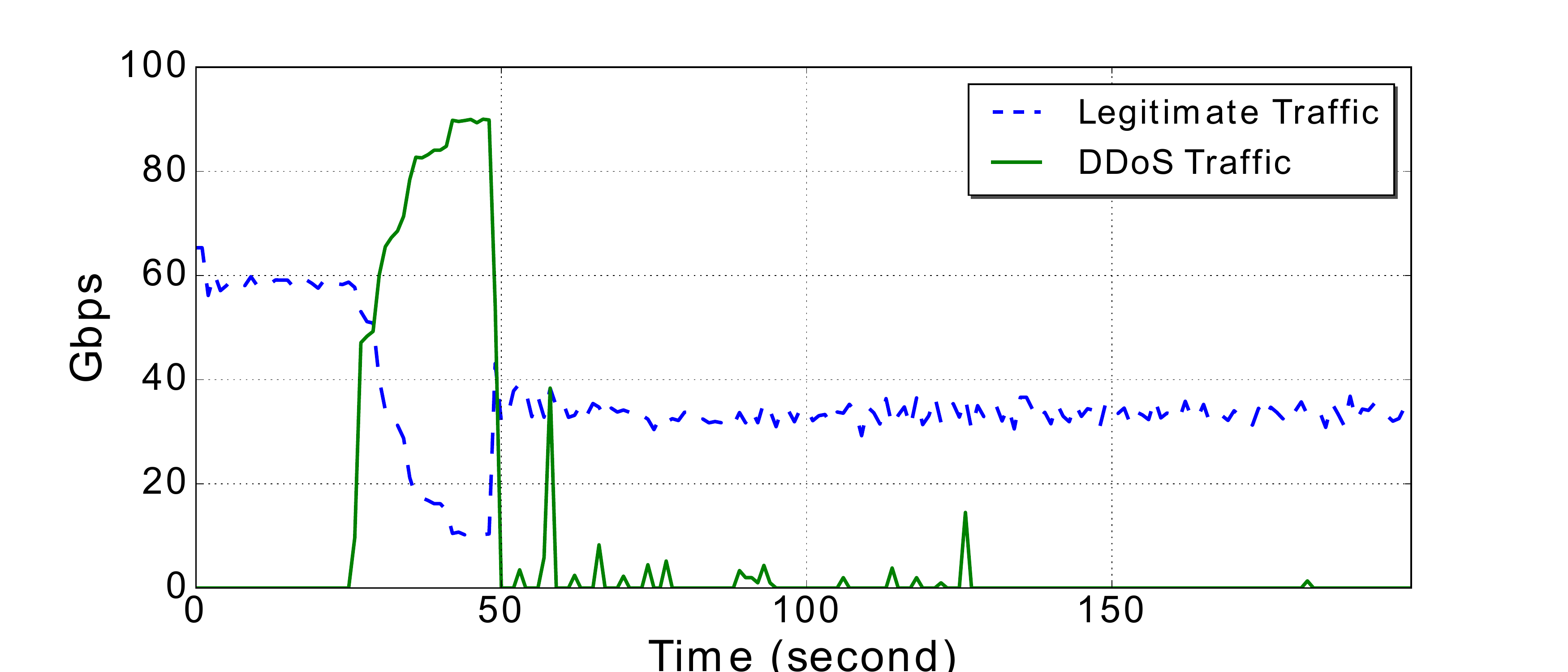}}
    \subfigure[Traffic time series under rules for minimal collateral damage] {
    \label{fig:geni-2}
    \includegraphics[width=0.49\textwidth]{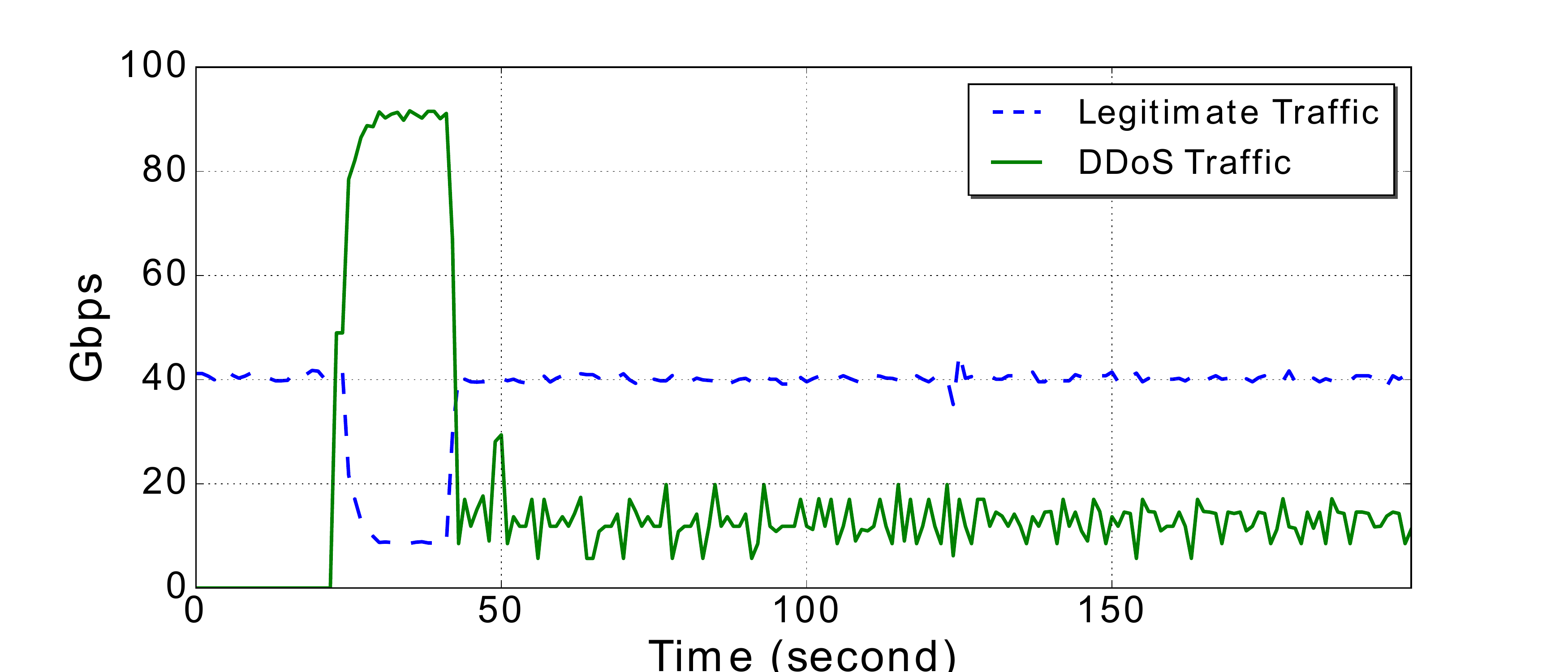}}
  \caption{\textbf{Volume of\consider{ incoming} legitimate and DDoS traffic over time before and during distributed filtering 
		of DDoS traffic during a pilot study.}}
  \label{fig:geni}
\end{figure}

Figure~\ref{fig:geni} shows our defense in two different scenarios.
In the first scenario (Figure~\ref{fig:geni-1}) where the defense begins at second 48,
it takes \emph{only about 3 seconds} for the filtering of DDoS traffic to reach 100\%.
Since we are using source-based filtering, 
and the number of attack sources (1000) is relatively high compared to the rule budget (150),
some collateral damage has to happen,
preventing the volume of legitimate traffic since second 48 from fully recovering;
nonetheless, relative to the sharp dip of DDoS traffic,
the legitimate traffic does recover to be between 30 and 40 Gbps.
In the second scenario (Figure~\ref{fig:geni-2}),
we increase the rule budget to 200 and minimize the collateral damage.
Although we no longer filter as much of the DDoS traffic as the first scenario,
we filter enough to relieve the link congestion,
while all the legitimate traffic can continue to flow at its previous rate.

Finally, we also conducted a pilot study with multiple major IXPs
to test the scalability of our system in the wild against real, large-scale DDoS attacks. 
The results are promising. 
For example, over a month at one IXP,
our system was able to generate rules towards minimal collateral damage
that covered 90\% of the attack traffic from all 46,552 attack IPs in less than 7 seconds. 

\section{Conclusion}
\label{sec:conclusion}

DDoS attacks are notorious for the damage they can cause to network users and services.
As a DDoS attack is to launch DDoS traffic from DDoS bots throughout the Internet towards a victim along many different paths,
this paper studies a distributed filtering model 
that allows nodes distributed along the paths of DDoS traffic to filter the DDoS traffic.
This model is made more feasible as nodes on the Internet 
are witnessing increased capabilities of inspecting and filtering network traffic.
This model is also necessary to avoid
not only the often futile filtering of traffic only at a couple convergence points,
but also the mistaken filtering of legitimate traffic 
as distributed filtering allows more filtering points including those only on the attack paths.

We focused on making distributed filtering adaptive.
For example, while traffic flows to filter may be defined by different source granularities,
we allow the filtering to adapt to the most effective granularity of each flow.
A subscriber
acting on behalf of a DDoS victim
can run a rule-generation algorithm to derive an $F$-tree
which, by tracking the traffic toward the victim and strategically aggregating flows from different sources,
can help derive traffic-filtering rules adaptive to different objectives 
and deploy the rules at the most effective filtering nodes. 

We thoroughly evaluated our system through large-scale simulations based on real-world DDoS attack traces,
including (i) the quality, quantity, and the tradeoff of rules generated with different objectives;
(ii) rule deployment success rates and distributions with different distributed, Internet-scale filtering profiles; and
(iii) the system's overall efficacy.
We also conducted pilot studies on a large-scale testbed as well as major IXPs,
showing our system can filter the DDoS traffic from large-scale DDoS attacks 
with as little as zero collateral damage
within several seconds.

\vspace{-0.5mm}
\section*{Acknowledgment}
\label{sec:ack}
This project is in part the result of funding provided by the Science and
Technology Directorate of the United States Department of Homeland Security
under contract number D15PC00204. The views and conclusions contained
herein are those of the authors and should not be interpreted necessarily
representing the official policies or endorsements, either expressed or implied,
of the Department of Homeland Security or the US Government.
\vspace{-0.6mm}

\bibliographystyle{IEEEtran}
\bibliography{main}

\begin{thebibliography}{10}
\providecommand{\url}[1]{#1}
\csname url@samestyle\endcsname
\providecommand{\newblock}{\relax}
\providecommand{\bibinfo}[2]{#2}
\providecommand{\BIBentrySTDinterwordspacing}{\spaceskip=0pt\relax}
\providecommand{\BIBentryALTinterwordstretchfactor}{4}
\providecommand{\BIBentryALTinterwordspacing}{\spaceskip=\fontdimen2\font plus
\BIBentryALTinterwordstretchfactor\fontdimen3\font minus
  \fontdimen4\font\relax}
\providecommand{\BIBforeignlanguage}[2]{{%
\expandafter\ifx\csname l@#1\endcsname\relax
\typeout{** WARNING: IEEEtran.bst: No hyphenation pattern has been}%
\typeout{** loaded for the language `#1'. Using the pattern for}%
\typeout{** the default language instead.}%
\else
\language=\csname l@#1\endcsname
\fi
#2}}
\providecommand{\BIBdecl}{\relax}
\BIBdecl

\bibitem{akamai2020report}
A.~Research, ``2020 state of the {I}nternet / security: 2020 – a year in
  review,''
  \url{https://www.akamai.com/us/en/multimedia/documents/state-of-the-internet/soti-security-a-year-in-review-report-2020.pdf},
  2020.

\bibitem{Dyn2016}
\BIBentryALTinterwordspacing
K.~York, ``Dyn's statement on the 10/21/2016 {DNS} {DDoS} attack,'' \emph{Dyn
  Blog}, 2016. [Online]. Available:
  \url{https://dyn.com/blog/dyn-statement-on-10212016-ddos-attack/}
\BIBentrySTDinterwordspacing

\bibitem{Github2018}
\BIBentryALTinterwordspacing
S.~Kottler, ``February 28th {DDoS} incident report,'' \emph{GitHub
  Engineering}, 2018. [Online]. Available:
  \url{https://githubengineering.com/ddos-incident-report/}
\BIBentrySTDinterwordspacing

\bibitem{microsoft2021attack}
S.~Vaughan-Nichols, ``Microsoft azure fends off huge {DDoS} attack,''
  \url{https://www.zdnet.com/article/microsoft-azure-fends-off-huge-ddos-attack/},
  2021.

\bibitem{microsoft2022attack}
A.~Toh, ``Azure ddos protection: 2021 q3 and q4 {DDoS} attack trends,''
  \url{https://azure.microsoft.com/en-us/blog/azure-ddos-protection-2021-q3-and-q4-ddos-attack-trends/},
  2022.

\bibitem{akamai2020ddos}
T.~Emmons, ``Largest ever recorded packet per second-based {DDoS} attack
  mitigated by {Akamai},''
  https://www.akamai.com/blog/news/largest-ever-recorded-packet-per-secondbased-ddos-attack-mitigated-by-akamai,
  2020.

\bibitem{rfc5575}
P.~Marques, N.~Sheth, R.~Raszuk, B.~Greene, J.~Mauch, and D.~McPherson, ``{RFC}
  5575: An infrastructure to support secure {I}nternet routing,'' 2009.

\bibitem{arboraps}
NETSCOUT, ``Arbor availability protection system,''
  \url{https://www.netscout.com/product/arbor-availability-protection-system},
  2019.

\bibitem{fastnetmon}
P.~Odintsov, ``{FastNetMon} community - very fast {DDoS} analyzer with {sFlow},
  {NetFlow}, mirror support,''
  \url{https://github.com/pavel-odintsov/fastnetmon}, 2019.

\bibitem{Posidon}
M.~Zhang, G.~Li, S.~Wang, C.~Liu, A.~Chen, H.~Hu, G.~Gu, Q.~Li, M.~Xu, and
  J.~Wu, ``Poseidon: Mitigating volumetric {DDoS} attacks with programmable
  switches,'' in \emph{The 27th Network and Distributed System Security
  Symposium (NDSS)}, 2020.

\bibitem{liu2021jaqen}
Z.~Liu, H.~Namkung, G.~Nikolaidis, J.~Lee, C.~Kim, X.~Jin, V.~Braverman, M.~Yu,
  and V.~Sekar, ``Jaqen: A high-performance switch-native approach for
  detecting and mitigating volumetric {DDoS} attacks with programmable
  switches,'' in \emph{{USENIX} Security Symposium}, 2021.

\bibitem{bohatei}
S.~K. Fayaz, Y.~Tobioka, V.~Sekar, and M.~Bailey, ``Bohatei: Flexible and
  elastic {DDoS} defense.'' in \emph{USENIX Security Symposium}, 2015.

\bibitem{akamaiscrubbing}
``Akamai {DDoS} protection service,''
  \url{http://www.akamai.com/us/en/solutions/
  products/cloud-security/ddos-protection-service.jsp}, Akamai Technologies,
  2016.

\bibitem{liu2016middlepolice}
Z.~Liu, H.~Jin, Y.-C. Hu, and M.~Bailey, ``Middlepolice: Toward enforcing
  destination-defined policies in the middle of the {Internet},'' in
  \emph{Proceedings of the 2016 ACM SIGSAC Conference on Computer and
  Communications Security}, ser. CCS, 2016.

\bibitem{mirkovic2002attacking}
J.~Mirkovic, G.~Prier, and P.~Reiher, ``Attacking {DDoS} at the source,'' in
  \emph{Network Protocols, 2002. Proceedings. 10th IEEE International
  Conference on}.\hskip 1em plus 0.5em minus 0.4em\relax IEEE, Nov. 2002, pp.
  312--321.

\bibitem{papadopoulos2003}
C.~Papadopoulos, R.~Lindell, J.~Mehringer, A.~Hussain, and R.~Govindan,
  ``{COSSACK: Coordinated Suppression of Simultaneous Attacks},'' \emph{DARPA
  Information Survivability Conference and Exposition}, vol.~2, 2003.

\bibitem{mahajan2002pushback}
R.~Mahajan, S.~M. Bellovin, S.~Floyd, J.~Ioannidis, V.~Paxson, and S.~Shenker,
  ``Controlling high bandwidth aggregates in the network,'' \emph{{SIGCOMM}
  Computer Communication Review}, vol.~32, no.~3, 2002.

\bibitem{yang2005TVA}
X.~Yang, D.~Wetherall, and T.~Anderson, ``{TVA}: A {DoS}-limiting network
  architecture,'' \emph{IEEE/ACM Transactions on Networking}, vol.~16, 2008.

\bibitem{Argyraki2005}
K.~Argyraki and D.~R. Cheriton, ``{Active Internet Traffic Filtering: Real-time
  Response to Denial-of-service Attacks},'' in \emph{USENIX Annual Technical
  Conference}, 2005.

\bibitem{oikonomou2006framework}
G.~Oikonomou, J.~Mirkovic, P.~Reiher, and M.~Robinson, ``A framework for a
  collaborative {DDoS} defense,'' in \emph{Computer Security Applications
  Conference (ACSAC)}, 2006, pp. 33--42.

\bibitem{liu2008filter}
X.~Liu, X.~Yang, and Y.~Lu, ``{To filter or to authorize: Network-layer DoS
  defense against multimillion-node botnets},'' \emph{ACM SIGCOMM Computer
  Communication Review}, 2008.

\bibitem{kline2009rad}
E.~Kline, M.~Beaumont-Gay, J.~Mirkovic, and P.~Reiher, ``{RAD}: Reflector
  attack defense using message authentication codes,'' in \emph{IEEE Annual
  Computer Security Applications Conference}, 2009.

\bibitem{ramanathan2018senss}
S.~Ramanathan, J.~Mirkovic, M.~Yu, and Y.~Zhang, ``{SENSS} against volumetric
  {DDoS} attacks,'' in \emph{Proceedings of the 34th Annual Computer Security
  Applications Conference}, ser. ACSAC, 2018.

\bibitem{dietzel2018stellar}
C.~Dietzel, M.~Wichtlhuber, G.~Smaragdakis, and A.~Feldmann, ``Stellar: Network
  attack mitigation using advanced blackholing,'' in \emph{Proceedings of the
  14th International Conference on Emerging Networking EXperiments and
  Technologies}, ser. CoNEXT, 2018.

\bibitem{savage2000practical}
S.~Savage, D.~Wetherall, A.~Karlin, and T.~Anderson, ``Practical network
  support for {IP} traceback,'' in \emph{ACM SIGCOMM}, 2000.

\bibitem{yaar2005fit}
A.~Yaar, A.~Perrig, and D.~Song, ``{FIT: fast {Internet} traceback},'' in
  \emph{Proceedings IEEE 24th Annual Joint Conference of the IEEE Computer and
  Communications Societies}, vol.~2, 2005, pp. 1395--1406.

\bibitem{argyraki2005active}
K.~J. Argyraki and D.~R. Cheriton, ``{Active Internet Traffic Filtering:
  Real-Time Response to Denial-of-Service Attacks},'' in \emph{USENIX annual
  technical conference, general track}, 2005, pp. 135--148.

\bibitem{snoeren2001hash}
A.~Snoeren, C.~Partridge, L.~Sanchez, C.~Jones, F.~Tchakountio, S.~Kent, and
  T.~Strayer, ``Hash-based {IP} traceback,'' in \emph{ACM SIGCOMM}, 2001.

\bibitem{li2004large}
J.~Li, M.~Sung, J.~Xu, and L.~Li, ``Large-scale {IP} traceback in high-speed
  {I}nternet: Practical techniques and theoretical foundation,'' in \emph{IEEE
  Symposium on Security and Privacy}, 2004, pp. 115--129.

\bibitem{shi22pathfinder}
L.~Shi, J.~Li, M.~Zhang, and P.~Reiher, ``On capturing {DDoS} traffic
  footprints on the {I}nternet,'' \emph{IEEE Transactions on Dependable and
  Secure Computing}, vol.~19, no.~4, pp. 2755--2770, 2022.

\bibitem{opendaylight2018}
OpenDayLight. (2018) Opendaylight platform overview.
  \url{https://www.opendaylight.org/what-we-do/odl-platform-overview}.

\bibitem{berde2014onos}
P.~Berde, M.~Gerola, J.~Hart, Y.~Higuchi, M.~Kobayashi, T.~Koide, B.~Lantz,
  B.~O'Connor, P.~Radoslavov, W.~Snow \emph{et~al.}, ``{ONOS}: Towards an open,
  distributed {SDN OS},'' in \emph{Proceedings of the third workshop on Hot
  topics in software defined networking}, 2014.

\bibitem{ryu2017}
Ryu. (2017) Ryu {SDN} framework. \url{https://osrg.github.io/ryu/}.

\bibitem{kang13crossfire}
M.~S. Kang, S.~B. Lee, and V.~D. Gligor, ``The crossfire attack,'' in
  \emph{IEEE Symposium on Security and Privacy}, 2013, pp. 127--141.

\bibitem{radb2016}
M.~Network, ``Merit {RAD}b,'' \url{http://www.radb.net}, 2016.

\bibitem{santannaBooters}
J.~Santanna, R.~van Rijswijk-Deij, R.~Hofstede, A.~Sperotto, M.~Wierbosch,
  L.~Zambenedetti~Granville, and A.~Pras, ``Booters - an analysis of
  {DDoS}-as-a-service attacks,'' in \emph{IFIP/IEEE International Symposium on
  Integrated Network Management (IM)}, 2015.

\bibitem{caida2007}
C.~for Applied Internet Data~Analysis. (2007) The {CAIDA} {UCSD} {DDoS} attack
  2007 dataset.
  \url{http://www.caida.org/data/passive/ddos-20070804_dataset.xml}.

\bibitem{GENI}
M.~Berman, J.~S. Chase, L.~Landweber, A.~Nakao, M.~Ott, D.~Raychaudhuri,
  R.~Ricci, and I.~Seskar, ``{GENI}: A federated testbed for innovative network
  experiments,'' \emph{Computer Networks}, vol.~61, pp. 5--23, 2014.

\bibitem{openvswitch}
B.~Pfaff, J.~Pettit, T.~Koponen, E.~Jackson, A.~Zhou, J.~Rajahalme, J.~Gross,
  A.~Wang, J.~Stringer, P.~Shelar, K.~Amidon, and M.~Casado, ``The design and
  implementation of {Open vSwitch},'' in \emph{12th {USENIX} Symposium on
  Networked Systems Design and Implementation ({NSDI})}, 2015.

\bibitem{ospf}
J.~Moy. (2017) {OSPF Version 2}. \url{https://tools.ietf.org/html/rfc2328}.

\end{thebibliography}

\end{document}